\documentclass{article}
\usepackage{bm,latexsym,amsmath,amsfonts,amssymb,fancyhdr,color,graphicx,multirow}
\usepackage[a4paper,bottom=3cm,top=3.cm,head=5mm,width=18cm,dvipdfm]{geometry}
\usepackage[usenames,dvipsnames,svgnames,table]{xcolor}
\usepackage[
            colorlinks=true,
            linkcolor=blue,
            urlcolor=blue,
            citecolor=green,          
            bookmarks=true,
            bookmarksnumbered=true,
            breaklinks=true,
            pdfpagemode=Fullscreen,
            pdfstartview=FitBH]{hyperref}
\usepackage[dotinlabels]{titletoc}
\usepackage{titlesec}

\newcommand{\titledef}{} 

\numberwithin{equation}{section}
\allowdisplaybreaks[4]

\titlelabel{\thetitle.\quad \hspace{-0.8em}}
\titlecontents{section}
              [1.5em]
              {\vspace{4mm} \large \bf}
              {\contentslabel{1em}}
              {\hspace*{-1em}}
              {\titlerule*[.5pc]{.}\contentspage}
\titlecontents{subsection}
              [3.5em]
              {\vspace{2mm}}
              {\contentslabel{1.8em}}
              {\hspace*{.3em}}
              {\titlerule*[.5pc]{.}\contentspage}
\titlecontents{subsubsection}
              [5.5em]
              {\vspace{2mm}}
              {\contentslabel{2.5em}}
              {\hspace*{.3em}}
              {\titlerule*[.5pc]{.}\contentspage}

\hypersetup{ pdfauthor = {Shao-Feng Ge},
	     pdftitle = {\titledef}, 
	     pdfsubject = {}, 
             pdfkeywords = {}, 
	     pdfcreator = {LaTeX with hyperref package},
	     pdfproducer = {dvips + ps2pdf} }

\definecolor{gesfpurple}{rgb}{0.47,0.19,0.42}

\definecolor{gesflanse}{rgb}{0.00,0.50,0.50}

\definecolor{gesfblue}{rgb}{0.08,0.42,0.76}
\newcommand{\gblue}[1]{{\color{gesfblue} #1}}
\definecolor{gesfred}{rgb}{1,0,0}
\newcommand{\gred}[1]{{\color{gesfred} #1}}
\definecolor{gesfwhite}{rgb}{1,1,1}

\definecolor{gesfblack}{rgb}{0,0,0}

\newcommand{\gsec}[1]{{\hypersetup{linkcolor=red}Sec.~\ref{#1}\hypersetup{linkcolor=blue}}}

\newcommand{\geqn}[1]{\hypersetup{linkcolor=blue}(\ref{#1})\hypersetup{linkcolor=blue}}
\newcommand{\gfig}[1]{{\hypersetup{linkcolor=violet}Fig.~\ref{#1}\hypersetup{linkcolor=blue}}}
\newcommand{\gtab}[1]{{\hypersetup{linkcolor=gesflanse}Tab.~\ref{#1}\hypersetup{linkcolor=blue}}}

\newcommand{\dMa}{\delta_{\rm M1}}
\newcommand{\dMb}{\delta_{\rm M2}}
\newcommand{\dMc}{\delta_{\rm M3}}
\newcommand{\dD}{\delta_D}
\newcommand{\Dma}{\Delta m^2_a}
\newcommand{\Dms}{\Delta m^2_s}
\newcommand{\mee}{\langle m \rangle_{ee}}
\newcommand{\nuless}{0 \nu 2 \beta}
\newcommand{\Tr}{\theta_r}
\newcommand{\Ts}{\theta_s}
\newcommand{\vLa}{\overrightarrow{L_1}}
\newcommand{\vLb}{\overrightarrow{L_2}}
\newcommand{\vLc}{\overrightarrow{L_3}}
\newcommand{\Thalf}{T^{0 \nu}_{1/2}}

\begin{document}
\fontsize{12pt}{14pt}\selectfont

\title{\textbf{\fontsize{13pt}{14pt}\selectfont Extracting Majorana Properties in the Throat of Neutrinoless Double Beta Decay}}
\author{{Shao-Feng Ge}\footnote{gesf02@gmail.com}\, and
        {Manfred Lindner}\footnote{lindner@mpi-hd.mpg.de} \\[3mm]
        {Max-Planck-Institut f\"{u}r Kernphysik, Heidelberg 69117, Germany}}
\date{\today}

\maketitle

\begin{abstract}
\fontsize{12pt}{14pt}\selectfont
Assuming that neutrinos are Majorana particles,
we explore what information can be inferred from future strong limits (i.e. non-observation)
for neutrinoless double beta decay. Specifically we consider the case where the
mass hierarchy is normal and the different contributions to the effective mass $\mee$ partly cancel.
We discuss how this fixes the two Majorana CP phases simultaneously from the Majorana Triangle and how it limits
the lightest neutrino mass $m_1$ within a narrow window. The two Majorana CP phases are
in this case even better determined than in the usual case for larger $\mee$.
We show that the
uncertainty in these predictions can be significantly reduced by the complementary measurement
of reactor neutrino experiments, especially the medium baseline version JUNO/RENO-50.
We also estimate the necessary precision on $\mee$ to infer
non-trivial Majorana CP phases and the upper limit $\mee \lesssim 1\,\mbox{meV}$ sets a target
for the design of future neutrinoless double beta decay experiments.
\end{abstract}

\vspace{1cm}

\newpage
\section{Introduction}

The neutrino has always been a mysterious particle since it was invented by Pauli in 1930s \cite{nuHistory}.
It only participates in weak interactions and is therefore difficult to detect.
Different from all other fermions, we have only observed the left-handed component of neutrinos.
In the Standard Model (SM) of particle physics \cite{SM}, the right-handed component and any
other operator that allows finite neutrino mass is absent.
The discovery of neutrino masses is therefore the first observation of some new physics (NP) beyond SM.
Equivalently, neutrino is massless in SM until neutrino oscillation \cite{PMNS}
is established is established by solar \cite{Solar} and atmospheric \cite{Atmos} experiments.
If neutrinos are massive, the oscillation phenomena can be explained by the non-trivial mixing
between different flavors. While neutrinos are produced and detected as flavor eigenstates in
association with charged leptons, they propagate as plane waves corresponding to mass
eigenstates. The tiny difference in the oscillation phases due to mass eigenvalues then
introduce coherent interference between neutrinos of different flavors.

Being a neutral fermion is another unique feature of the neutrino. It can be either a Dirac or Majorana
type fermion \cite{Majorana:1937vz}. Correspondingly, it can have either a Dirac mass term, which
connects the left- and right-handed components, or a Majorana mass term, which involves only
left-handed components \cite{Willenbrock:2004hu}.
While the Dirac mass term conserves lepton number, Majorana mass term violates it. To explain
neutrino masses, either right-handed components must exist to allow Dirac masses or there is
lepton number violation \cite{Duerr:2011zd} to produce Majorana masses. Either way,
the SM needs to be extended to incorporate new physics.

The difference between Dirac and Majorana mass terms affects processes involving an
intermediate neutrino propagator. A perfect testing ground is neutrinoless double beta
($\nuless$) decay \cite{nuless}, $(A,Z) \rightarrow (A, Z+2) + 2 e^-$, where the
nuclei $(A,Z)$ decays into $(A,Z)$ with two electrons, and no neutrino in the final state.
The half-lifetime ($\Thalf$) is inversely proportional to the effective mass $\mee^2$, with the
subscript $ee$ denoting the two final-state electrons. Although there are other types of process that
can manifest the Majorana nature of light neutrinos, such as neutrino-antineutrino oscillation
\cite{nu-antinu} or inverse neutrinoless double beta decay \cite{nu-collider},
$\nuless$ decay is the most promising process under pursuit \cite{nuless-review}.
Observing $\nuless$ decay would establish lepton number violation which could entirely be due to
Majorana masses. The observation implies also a Majorana component of light neutrinos \cite{SchechterValle}, 
but it could also point to some other lepton number violation which induces only an extremely tiny
Majorana component \cite{Duerr:2011zd}.

Currently, there are many experimental searches for this rare process of $\nuless$ decay
\footnote{We list all experiments (existing and those in the future) here and present the
current best 90\% limits on the half-lifetime $\Thalf$. Please check \cite{rev16} for more details.}.
Mainly five elements ($^{130}$Te, $^{76}$Ge, $^{100}$Mo, $^{136}$Xe, and $^{82}$Se) have
been used as target material.
1) Cuoricino \cite{Cuoricino}, CUORE \cite{Cuore0, Cuore} and SNO+ \cite{SNO+} use $^{130}\mbox{Te}$
with the current best limit $\Thalf \geq 2.9 \times 10^{24}$\,yr from CUORE-0 \cite{Cuore0}.
2) $^{76}$Ge has been used by five experiments: Heidelberg-Moscow
\cite{Heidelberg-Moscow}, IGEX \cite{IGEX}, GERDA-I \cite{GERDA-I}, GERDA-II \cite{GERDA-II},
and Majorana Demonstrator \cite{MajoranaD}, of which GERDA-II has the best limit
$\Thalf \geq 5.2 \times 10^{25}$\,yr \cite{Gerda-nu16}.
There are plans to use $^{76}$Ge for upgrades in $\mathcal O$(200kg) experiments or new ton-scale detectors.
3) $^{136}$Xe is used in the current experiments EXO-200 \cite{EXO} and KamLAND-Zen \cite{KamLAND-Zen}
with best limit
$\Thalf \geq 1.1 \times 10^{26}$\,yr from the latter. The future experiments NEXT \cite{NEXT},
nEXO \cite{nEXO}, and PandaX-III \cite{PandaX-III} also use $^{136}$Xe as experiment material.
4) $^{100}$Mo has been used in NEMO-3 \cite{NEMO-3} to obtain 
$\Thalf \geq 1.1 \times 10^{24}$\,yr and will be used in AMoRE \cite{AMoRE}.
5) For $^{82}$Se, it has not be used ever yet but has already been chosen by
LUCIFER \cite{LUCIFER} and SuperNEMO \cite{SuperNEMO}.

$\nuless$ decay has so far not been observed \cite{rev16}.
The effective mass $\mee$ and hence $\nuless$ decay could even vanish
\cite{Bilenky:2001rz, Rodejohann:2002ng, Xing03} for the normal hierarchy (NH)
which is already somewhat preferred by both cosmological constraint \cite{NH:cosmology} and
the latest global fit of neutrino oscillation \cite{NH:osc}.
There are two Majorana CP phases providing enough degrees of freedom
for tiny $\nuless$ decay, and we will discuss that vanishing
$\mee$ can uniquely fix the two Majorana CP phases simultaneously. In the sense
of fixing the free parameters of $\nuless$ decay, including both Majorana CP phases
and the absolute mass scale, non-observation is even better.

We first use current measurement of neutrino oscillation parameters
and the cosmological constraint on the neutrino mass sum to predict the probability
distribution of the effective mass $\mee$ in \gsec{sec:mee} to show that non-observation
of $\nuless$ decay at next-generation experiments has sizable probability to happen.
This motivates our exploration in \gsec{sec:triangle} how vanishing $\mee$ can determine the two Majorana
CP phases with geometrical argument. Then we study the uncertainty from neutrino oscillation
parameters and point out the improvement from the future medium baseline reactor neutrino
experiments JUNO/RENO-50 in \gsec{sec:uncertainty}. To guarantee the extraction of non-trivial
Majorana CP phases puts stringent requirement on the future $\nuless$ decay experiments and
we study this quantitatively in \gsec{sec:sensitivity}. Our conclusions can be found in
\gsec{sec:conclusion}.

\section{The Effective Mass $\mee$ Under Current Prior Knowledge}
\label{sec:mee}

The neutrino mixing between flavor and mass eigenstates, $\nu_\alpha = U_{\alpha i} \nu_i$
($\alpha = e, \mu, \tau$ for flavor and $i = 1,2,3$ for mass) can be parametrized as,
\begin{equation}
  U
=
  \mathcal P
\left\lgroup
\begin{array}{ccc} 
  c_s c_r & s_s c_r & s_r e^{- i \dD} \\
- c_a s_s - s_a s_r c_s e^{i \dD} & c_a c_s - s_a s_r s_s e^{i \dD} & s_a c_r \\
  s_a s_s - c_a s_r c_s e^{i \dD} &-s_a c_s - c_a s_r s_s e^{i \dD} & c_a c_r
\end{array}
\right\rgroup
  \mathcal Q \,.
\label{eq:U}
\end{equation}
For convenience, we denote the three mixing angles and the two mass splits as,
\begin{equation}
  \theta_a \equiv \theta_{23} \,,
\qquad
  \theta_r \equiv \theta_{13} \,,
\qquad
  \theta_s \equiv \theta_{12} \,,
\qquad
  \Delta m^2_a \equiv \Delta m^2_{13} \,,
\qquad
  \Delta m^2_s \equiv \Delta m^2_{12} \,,
\end{equation}
according to the major processes through which these parameters are measured. The matrices
$\mathcal P \equiv \mbox{diag}\{e^{- i \beta_1}, e^{- i \beta_2}, e^{- i \beta_3}\}$ and
$\mathcal Q \equiv \mbox{diag}\{e^{- i \dMa/2}, e^{- i \dMb/2}, e^{- i (\dMc - \dD)/2}\}$
on the two sides are diagonal rephasing matrices. While the three phases $\beta_i$ in
$\mathcal P$ are unphysical, $\mathcal Q$ contains
two independent Majorana CP phases. In this paper we take $\dMb=0$ and
parametrize $\dMc$ in association with the Dirac CP phase $\dD$ for simplicity.
Then, only $\dMa$ and $\dMc$ would appear in the effective mass 
$\mee \equiv \sum_i m_i U^2_{ei}$,
\begin{equation}
  \mee
=
  m_1 |U_{e1}|^2 e^{i \dMa}
+ m_2 |U_{e2}|^2
+ m_3 |U_{e3}|^2 e^{i \dMc} \,,
\end{equation}
for $\nuless$ decay. The discussion on the two Majorana CP phases then decouples from the
unknown Dirac CP phase $\dD$. For normal hierarchy, the effective mass $\mee$ becomes,
\begin{equation}
  \mee
=
  m_1 c^2_r c^2_s e^{i \dMa}
+ \sqrt{m^2_1 + \Delta m^2_s} c^2_r s^2_s
+ \sqrt{m^2_1 + \Delta m^2_a} s^2_r e^{i \dMc} \,.
\end{equation}

The effective mass $\mee$ involves 7 independent parameters. Four of them, $\Tr$, $\Ts$, $\Dma$,
and $\Dms$, have been constrained by neutrino oscillation experiments. Across this paper, our input
\begin{subequations}
\begin{eqnarray}
  \Tr
=
  8.5^\circ \pm 0.2^\circ \,,
&&
  \Dma
=
  (2.457 \pm 0.047) \times 10^{-3} \mbox{eV}^2 \,,
\\
  \Ts
=
  33.48^\circ \pm 0.76^\circ \,,
&&
  \Dms
=
  (7.50 \pm 0.18) \times 10^{-5} \mbox{eV}^2 \,,
\end{eqnarray}
\label{eq:osc-inputs}
\end{subequations}
for NH is adopted according to the global fit \cite{globalfit15}.
We can produce a distribution of $\mee$ as a function of $m_1$
by sampling the four oscillation parameters according to \geqn{eq:osc-inputs}
and the two Majorana CP phases ($\dMa$ and $\dMc$) uniformly within $[0, 2\pi]$.
\begin{figure}[h]
\centering
\includegraphics[width=0.8\textwidth, height=10cm]{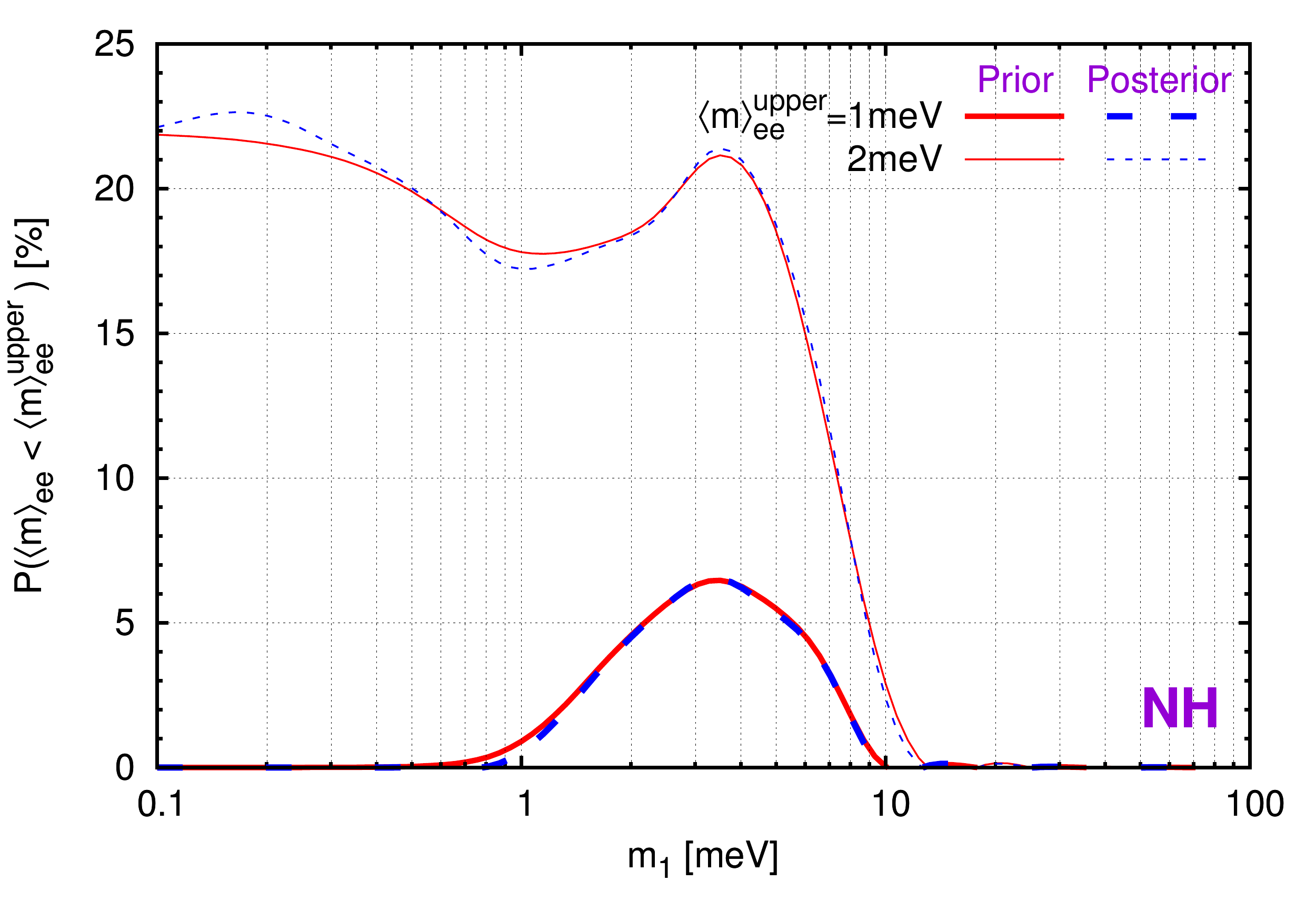}
\caption{The probability of $\mee < 1\,\mbox{meV}$ (thick) and  $\mee < 2\,\mbox{meV}$ (thin) for NH and given value of $m_1$,
         before (prior as solid lines) and after (posterior as dashed lines) JUNO/RENO-50
         experiments.}
\label{fig:probMee}
\end{figure}
In \gfig{fig:probMee}, we show the probability of $\mee$ being below $1\,\mbox{meV}$
and $2\,\mbox{meV}$ for NH, as a function of $m_1$. For $1\,\mbox{meV} \lesssim m_1 \lesssim 10\,\mbox{meV}$,
the effective mass $\mee$ has as large as $7\%$ of chance to be smaller than $1\,\mbox{meV}$
\cite{Benato:2015via}. Above $\mee^{upper} = 1\,\mbox{meV}$, the chance increases very fast.
For $\mee^{upper} = 2\,\mbox{meV}$, the chance jumps to around 20\% once
$m_1$ goes below $10\,\mbox{meV}$.
We show the results before and after JUNO/RENO-50 as solid and dashed lines for comparison.
Although the precision measurement of the solar angle $\Ts$ has significant effect on the
lower limit of $\mee$ for both NH and IH \cite{Ge:2015bfa}, its effect on the probability
$P(\mee < \mee^{upper})$ is not that significant after marginalization.

Recently, the cosmological data provide the most stringent constraint on the scale of neutrino
masses \cite{NH:cosmology} preferring slightly NH. Since the two mass squares
$\Dma$ and $\Dms$ have been measured, the cosmological data can also constrain
the lightest mass $m_1$ \cite{m1:cosmology}.
\begin{figure}[h]
\centering
\includegraphics[width=0.48\textwidth]{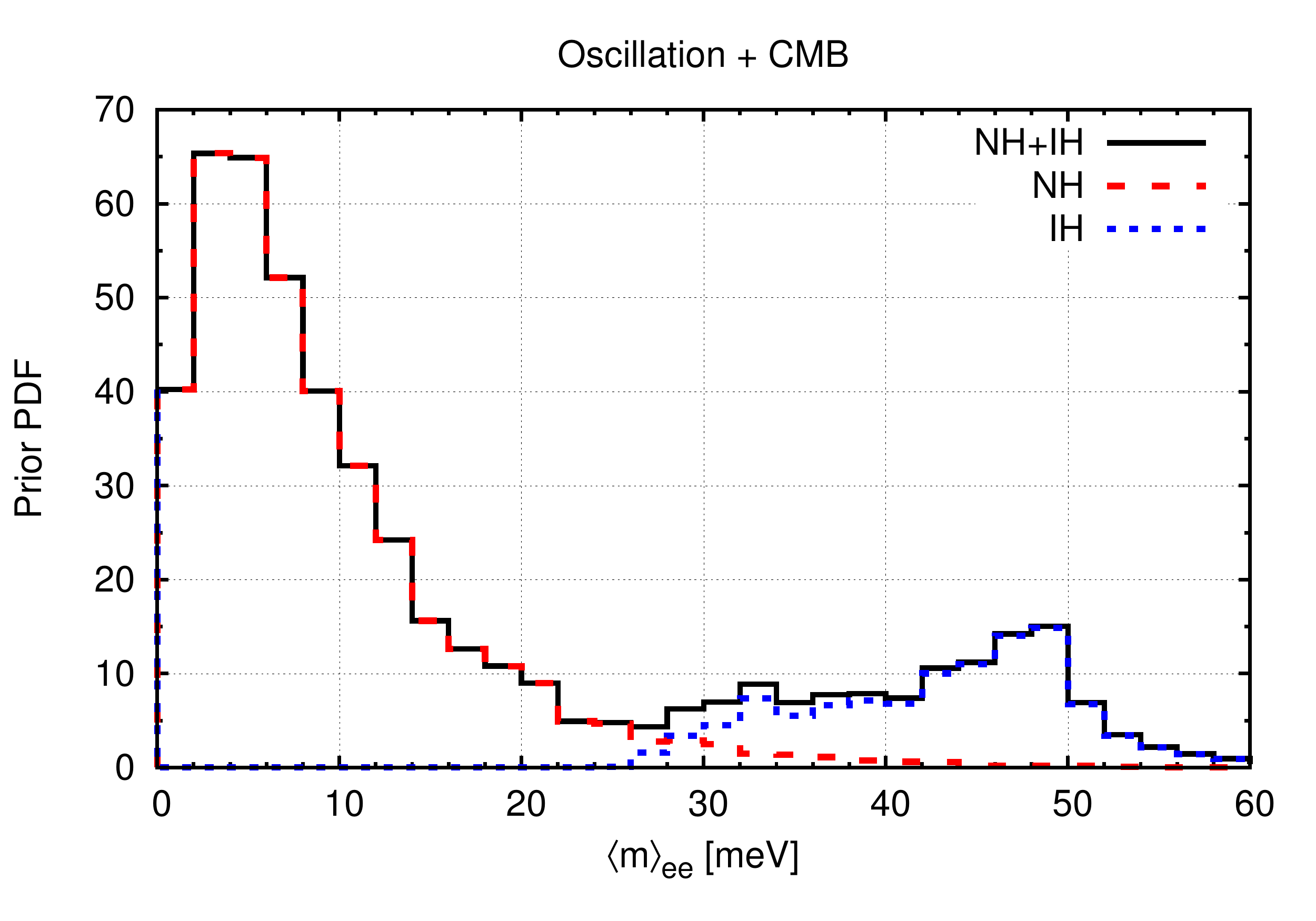}
\includegraphics[width=0.48\textwidth]{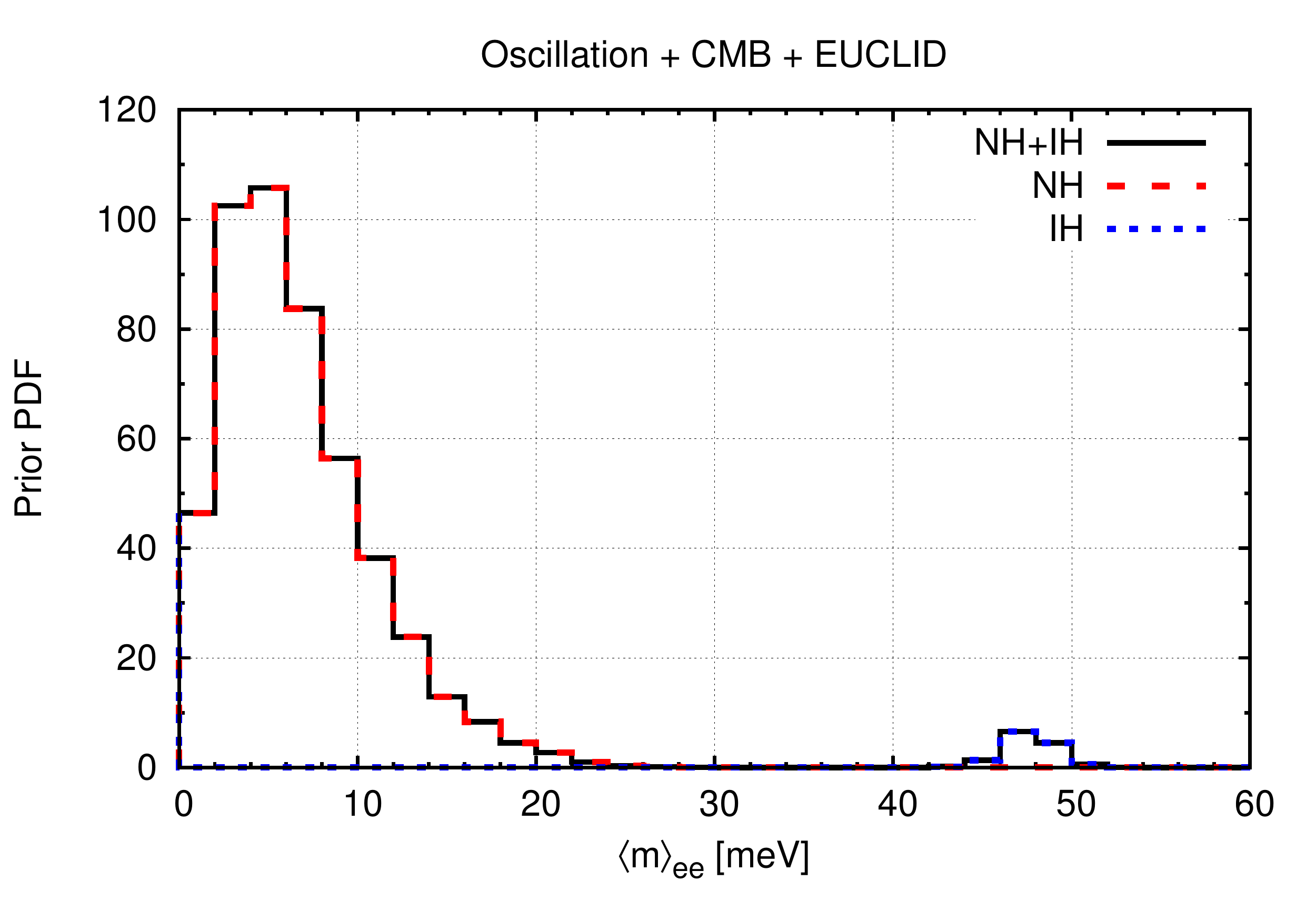}
\caption{The probability distribution function (PDF) of the $\nuless$ decay effective mass $\mee$,
         sampled from the global fit \cite{globalfit15} of the neutrino oscillation
         parameters (oscillation), and the constraint on $m_1$ from the Planck data (CMB)
         plus the prospective data from EUCLID-like survey (EUCLID) \cite{m1:cosmology}.
         Both NH (long-dashed red line) and
         IH (short-dashed blue line) have been sampled. The combined effect of NH
         and IH is shown as solid black line.}
\label{fig:mee}
\end{figure}
In \gfig{fig:mee}, we show the sampled distribution of $\mee$ from both neutrino oscillation
measurements \cite{globalfit15} and cosmological constraint \cite{m1:cosmology}.
The cosmological data predicts the probability for NH versus IH to be
around 2:1. This appears in the left panel of \gfig{fig:mee} as a larger peak around
$\mee \approx 5\,\mbox{meV}$ for NH while a smaller peak around $\mee \approx 50\,\mbox{meV}$
for IH. It can further increase to 12:1 if the prospective
observation from a EUCLID-like survey is added. Then, the IH peak in the $\mee$ distribution
in the right panel of \gfig{fig:mee} almost vanishes. The whole picture would not
be affected much by precision measurement at future medium baseline reactor neutrino experiments
JUNO/RENO-50.
\begin{figure}[h]
\centering
\includegraphics[width=0.8\textwidth, height=10cm]{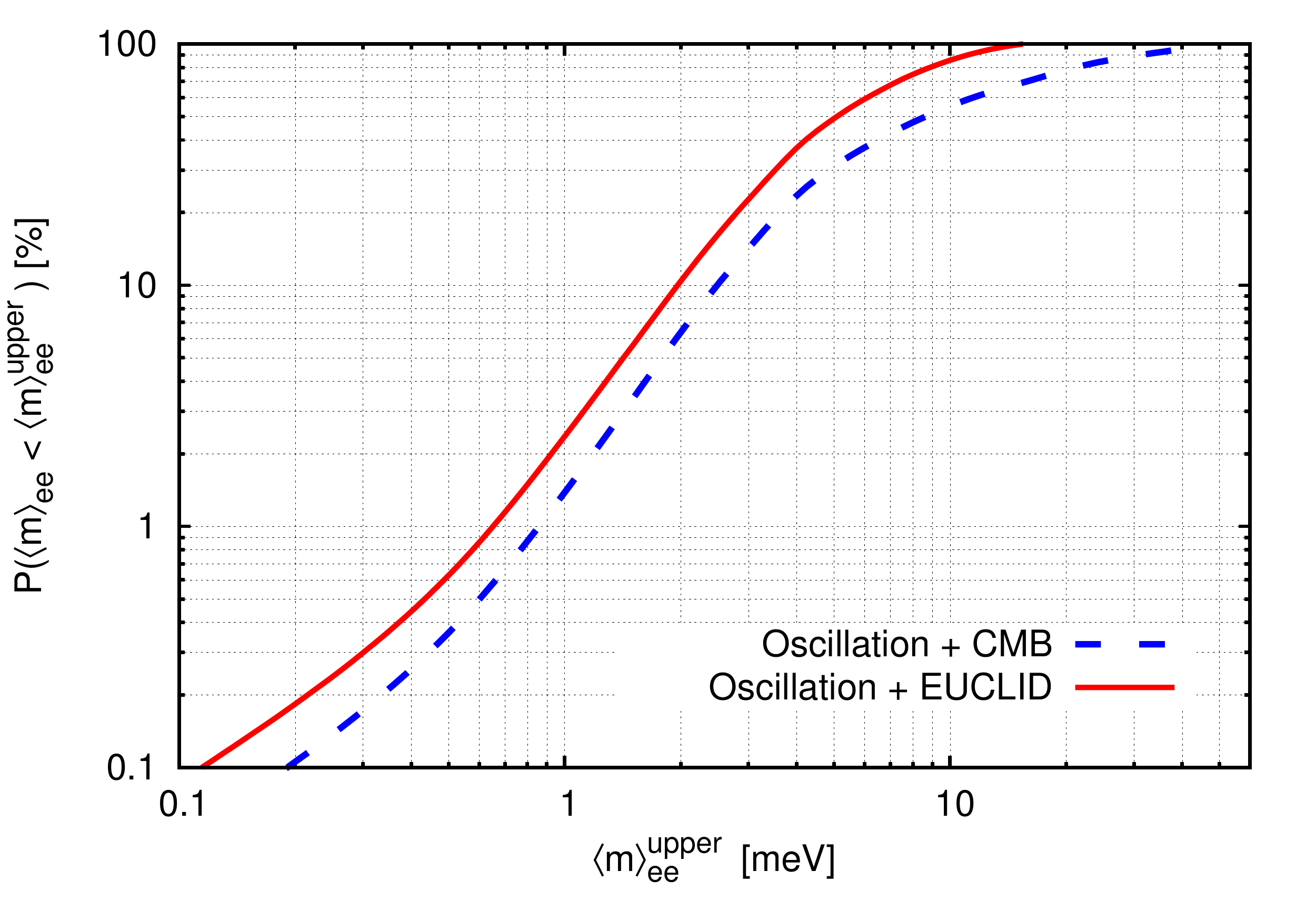}
\caption{The predicted probability of $\mee < \mee^{upper}$ with neutrino oscillation measurements
and Planck data (Oscillation + CMB) or the prospective observation from EUCLID-like
survey (Oscillation + EUCLID), as a function of $\mee^{upper}$.}
\label{fig:probMeeM0}
\end{figure}

To show the picture more clearly, we plot the probability of $\mee < \mee^{upper}$ as a function
of the upper value $\mee^{upper}$ in \gfig{fig:probMeeM0}, after folding the cosmological data with
the measurement of neutrino oscillation experiments. The curve starts from $P(\mee < 0) = 0\%$
to $100\%$ at large enough $\mee^{upper} \sim \mathcal O(10)\,\mbox{meV}$.
Note that the global lower limit of $\mee$
for IH is around $13\,\mbox{meV}$ \cite{Ge:2015bfa} and the chance for
$\mee \lesssim 10\,\mbox{meV}$ is quite close to the naive estimation $67\%$ (92\%) from
the probability ratio $P(NH):P(IH) \approx 2:1$ (12:1).
For more stringent constraints, $\mee$ has 1.3\% (6\%) of
chance to be smaller than $1\,\mbox{meV}$ ($2\,\mbox{meV}$). It significantly increases to
$2.2\%$ ($10\%$) if the EUCLID survey is available.

From the current constraints, the effective mass $\mee$ has a sizable chance to fall into the
throat of the NH chimney which would imply a non-observation
at current and up-coming $\nuless$ decay experiments.
Assuming that neutrinos are Majorana particles, we can then extract
from the non-observation interesting results.

\section{Extracting Majorana CP Phases from the Majorana Triangle}
\label{sec:triangle}

A non-observation of $\nuless$ decay does not exclude the possibility of Majorana neutrinos.
Since the $\nuless$ signal is proportional to the $\mee$, it is possible that the Majorana
CP phases $\dMa$ and $\dMc$ are such that there is no signal in the $ee$ channel. Reversely,
non-observation can pin down $\dMa$ and $\dMc$ under the condition of neutrinos are Majorana
particles.

\begin{figure}[h]
\centering
\includegraphics[width=8cm]{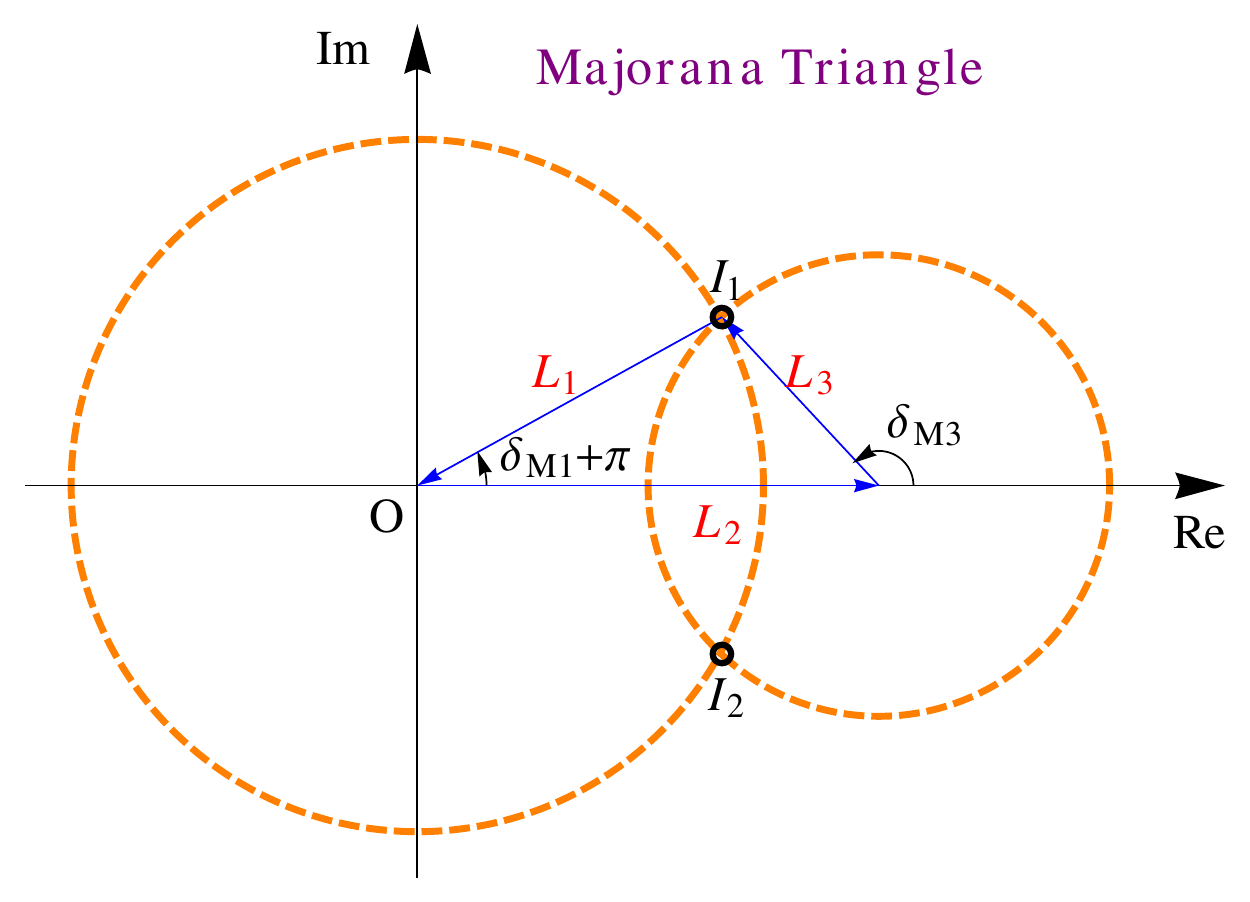}
\caption{The Majorana Triangle in $\nuless$ decay with vanishing $\mee$.}
\label{fig:mee1}
\end{figure}
For illustration, we adopt the geometric plot \cite{Xing:2014yka} which is a variant of
the Vissani graph \cite{Vissani:1999tu}. In the complex plane, $\mee$ is a vector sum,
\begin{equation}
  \mee
\equiv
  \vLa + \vLb + \vLc \,,
\end{equation}
as shown in \gfig{fig:mee1}. The three sides of the triangle are defined as,
\begin{subequations}
\begin{eqnarray}
&&
  \vLa \equiv m_1 U^2_{e1} = m_1 c^2_r c^2_s e^{i \dMa} \,,
\\
&&
  \vLb \equiv m_2 U^2_{e2} = \sqrt{m^2_1 + \Delta m^2_s} c^2_r s^2_s \,,
\\
&&
  \vLc \equiv m_3 U^2_{e3} = \sqrt{m^2_1 + \Delta m^2_a} s^2_r e^{i \dMc} \,.
\end{eqnarray}
\end{subequations}
Correspondingly, the length of the
three sides, $L_1 = m_1 c^2_r c^2_s$, $L_2 = m_2 c^2_r s^2_s$, and
$L_3 = m_3 s^2_r$, is modulated by $m_1$, $m_2$, and $m_3$, respectively.
In principle, there are three Majorana CP phases and only the two differences between them are physical.
In the Vassani graph, $\dMa$ is taken to be zero and $\vLa$ lies along the x-axis.
This choice is convenient for vanishing $m_1$. Nevertheless, vanishing $\mee$ can
only happen for nonzero $m_1$ with normal hierarchy. For this case, it is equivalent to
take any one of three Majorana CP phases to be zero. With vanishing $\dMb$, $\vLb$ lies along
the x-axis while the other two vectors $\vLa$ and $\vLc$ rotate around the two ends of
$\vLb$. Varying the two Majorana CP phases $\dMa$ and $\dMc$
\footnote{which are actually $\dMa - \dMb$ and $\dMc - \dMb$, respectively, with vanishing $\dMb$.},
namely the direction of $\vLa$ and $\vLc$, draws two circles on the complex plane.
The effective mass $\mee$ is then the vector between two arbitrary points
on the two circles, respectively.

\begin{figure}[h]
\centering
\includegraphics[width=0.45\textwidth]{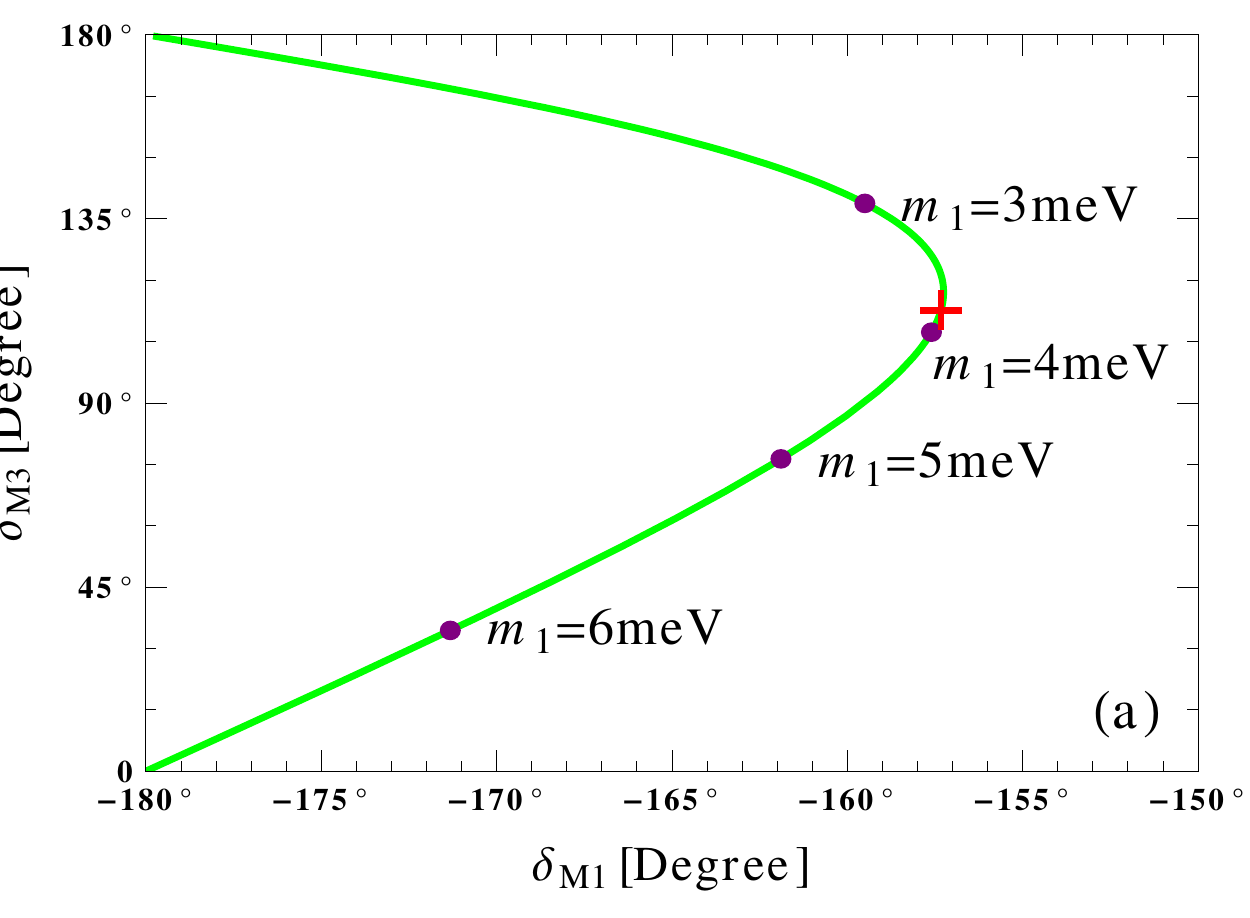}
\qquad
\includegraphics[width=0.45\textwidth]{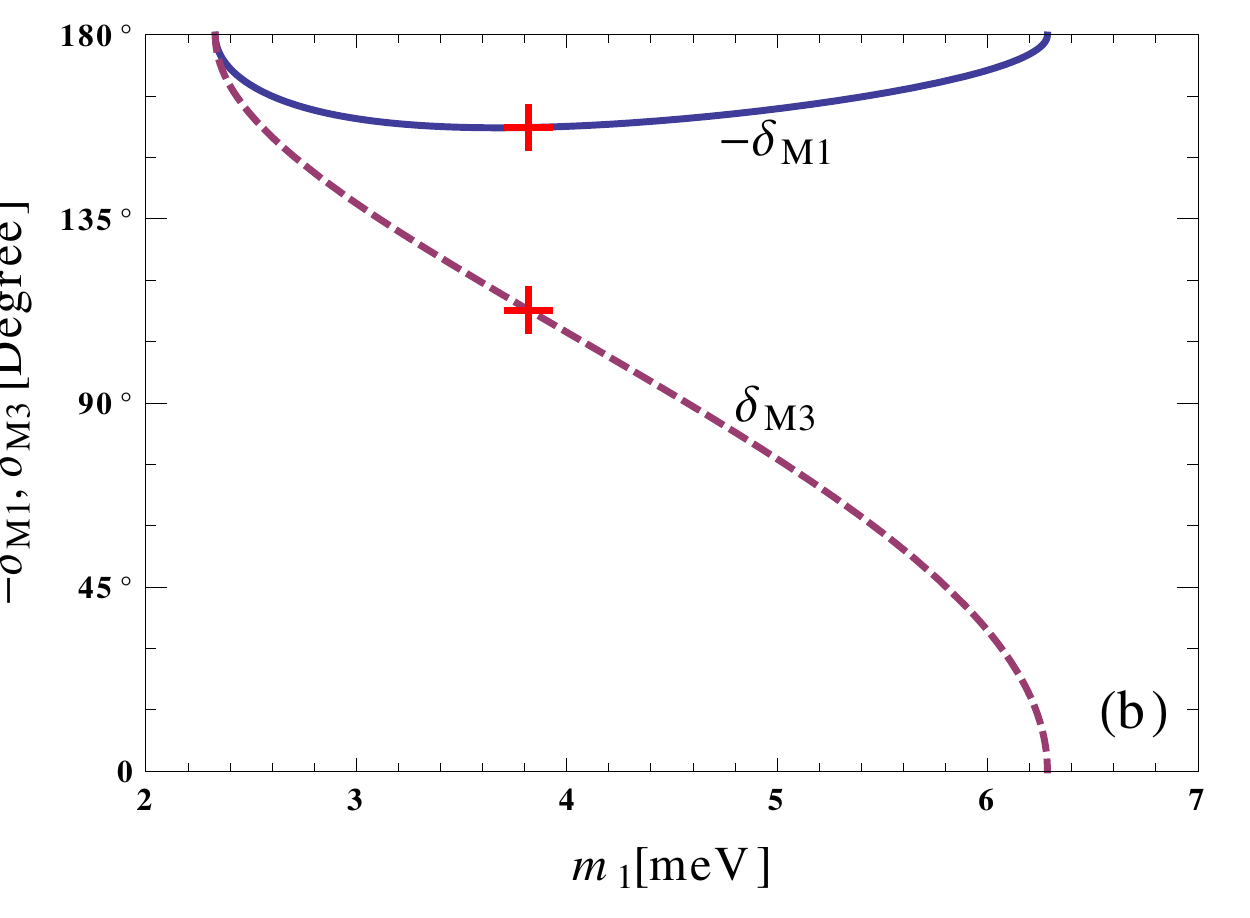}
\caption{The predicted Majorana CP phases from vanishing $\mee$ with both (a) two-dimensional
         plot $\dMa$--$\dMc$ and (b) one-dimensional $\dMa(m_1)$/$\dMc(m_1)$ as 
         implicit and explicit functions of the smallest mass $m_1$.}
\label{fig:dCPM13}
\end{figure}

As shown in \gfig{fig:mee1}, the three sides ($\vLa$, $\vLb$, $\vLc$) can form a {\it Majorana Triangle}
with vanishing $\mee$ if the two circles touch each other \cite{Xing:2014yka},
\begin{equation}
\left| L_1 - L_3 \right|
\leq
  L_2
\leq
  L_1 + L_3 \,.
\label{eq:conditions}
\end{equation}
It can happen at two intersection points, $I_1$ and $I_2$ as shown in
\gfig{fig:mee1}. Different from quadrilateral, the sides and angles of a triangle
has unique correlation with each other. From the length of the three sides,
we can immediately solve the two Majorana CP phases \footnote{For comparison, one
of the Majorana CP phases $\rho$ ($\equiv \dMa$) is also obtained as a function of
the smallest mass $m_1$ \cite{Xing:2015zha}.},
\begin{subequations}
\begin{eqnarray}
  \cos \dMa
& = &
- \frac {L^2_1 + L^2_2 - L^2_3}
				{2 L_1 L_2}
=
- \frac {m^2_1 c^4_r c^4_s + m^2_2 c^4_r s^4_s - m^2_3 s^4_r}
				{2 m_1 m_2 c^4_r c^2_s s^2_s} \,,
\\
  \cos \dMc
& = &
+ \frac {L^2_1 - L^2_2 - L^2_3}
				{2 L_2 L_3}
=
+ \frac {m^2_1 c^4_r c^4_s - m^2_2 c^4_r s^4_s - m^2_3 s^4_r}
				{2 m_2 m_3 c^2_r s^2_r s^2_s} \,.
\end{eqnarray}
\label{eq:cos}
\end{subequations}
The length of the three sides ($L_1$, $L_2$, $L_3$) are functions of oscillation
parameters ($\Dma$, $\Dms$, $\Tr$, $\Ts$) and the absolute mass scale $m_1$. Most of them
can be measured by neutrino oscillation experiments while the mass scale $m_1$ remains
a free parameter. With all oscillation parameters fixed, the vanishing $\mee$ would draw a line
in the two-dimensional space of $\dMa$ and $\dMc$, as implicit functions of
the mass scale $m_1$ shown in \gfig{fig:dCPM13}(a). For comparison, we also show the explicit
functions of $\dMa(m_1)$ and $\dMc(m_1)$ in \gfig{fig:dCPM13}(b). The cosine functions \geqn{eq:cos}
have two solutions, one in the upper complex plane and the other in the lower plane.
Due to symmetry, both solutions can exist, but
for simplicity, we show only one of them.
Note that $\dMa(m_1)$ and $\dMc(m_1)$ always appear in opposite planes. To be consistent
with the \gfig{fig:mee1}, we show the solution with $-180^\circ \leq \dMa \leq 0^\circ$
and $0^\circ \leq \dMc \leq 180^\circ$.

Across the interested region, \geqn{eq:conditions} or equivalently
$2.3\,\mbox{meV} \lesssim m_1 \lesssim 6.3\,\mbox{meV}$, as will be elaborated in \gsec{sec:sensitivity}
and shown in \gfig{fig:meeUpLimit}, $L_1$ increases linearly with $m_1$ while $L_3$ almost remains the same.
In addition, $L_1$ is always larger than $L_3$. Although $L_1$ is proportional to the smallest mass $m_1$
while $L_3$ is proportional to the much larger $m_3$, there is an extra suppression $s^2_r \approx 2.3\%$
associated with $m_3$. Consequently, the intersection points $I_1$ and $I_2$ are
always on the right-hand side of the origin $O$, see \gfig{fig:mee1}. Further,
the vector $\vLc$ can take any direction since $I_1$ and $I_2$ can take any point of the
smaller circle. Correspondingly, the $\vLa$ circle in \gfig{fig:mee1} expands with $m_1$, first
approaches the $\vLc$ circle with almost constant radius from the left, crosses it when
$L_1 = L_2 - L_3$, and finally swallows it when $L_1 = L_2 + L_3$. In this process,
$\dMc(m_1)$ decreases from $180^\circ$ to $0^\circ$. On the other hand, $\dMa(m_1)$ first
increases from $-180^\circ$ to its maximal value when the three sides form a right triangle,
$L^2_2 = L^2_1 + L^2_3$, and then decreases back to $-180^\circ$. The turning point
happens around,
\begin{equation}
  m^2_1
=
  \frac {c^4_r s^4_s \Dms - s^4_r \Dma}
				{c^4_r (c^2_s - s^2_s) + s^4_r}
\approx
  \frac {s^4_s}{c^2_s - s^2_s} \Dms \,.
\label{eq:turningPoint}
\end{equation}
Since $s^2_r \sim \mathcal O(\Dms/\Dma)$, the numerator is dominated by $c^4_r s^4_s \Dms$
while the denominator mainly comes from $c^4_r (c^2_s - s^2_s)$. Note that the omitted
contributions are introduced by $L_3$. The turning point roughly corresponds to $L_1 = L_2$
and is dictated by the solar parameters $\Ts$ and $\Dms$. Taking the current best fit
values, the turning point happens around $m_1 = 4.3\,\mbox{meV}$ with 
$\dMa = -158^\circ$ and $\dMc = 112^\circ$. To make it explicit,
the turning point has been shown in \gfig{fig:dCPM13} as red crosses. Although
\geqn{eq:turningPoint} is based on the observation that $L_1 > L_3$ and $L_3$
remains approximately constant, it approximates the turning points very precisely.
Since the three sides form a right triangle, the two Majorana CP phases are
correlated with each other, $\dMc = 270^\circ + \dMa$, at the turning point.

\section{Uncertainties and Improvement from Reactor Neutrino Experiments}
\label{sec:uncertainty}

Considering the fact that the oscillation parameters ($\Dma$, $\Dms$, $\Tr$, $\Ts$) are
not exactly measured, the prediction of $\dMa$ and $\dMc$ from the Majorana Triangle
would become a band, instead of the single line in \gfig{fig:dCPM13}\,(a).
\begin{figure}[h!]
\centering
\includegraphics[width=0.4\textwidth]{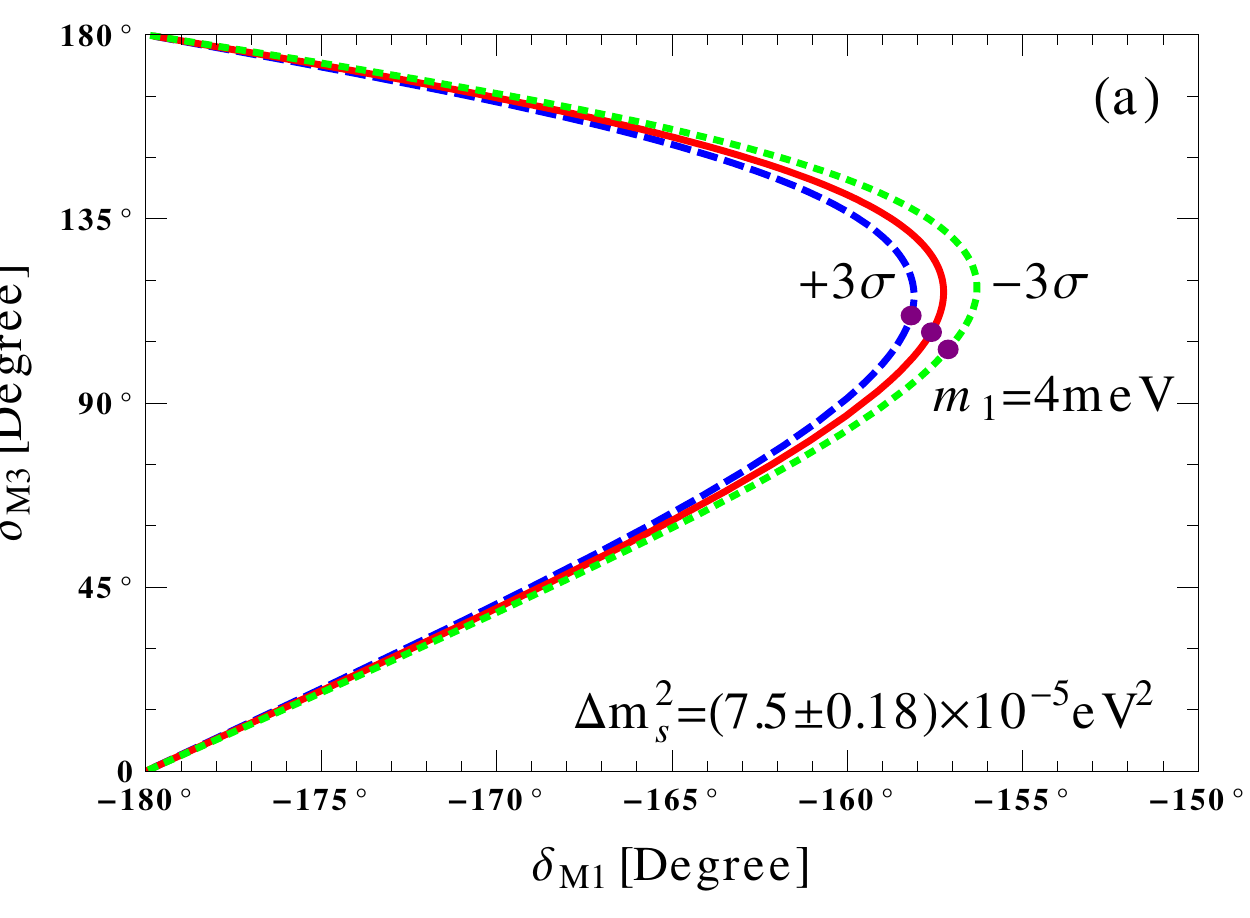}
\includegraphics[width=0.4\textwidth]{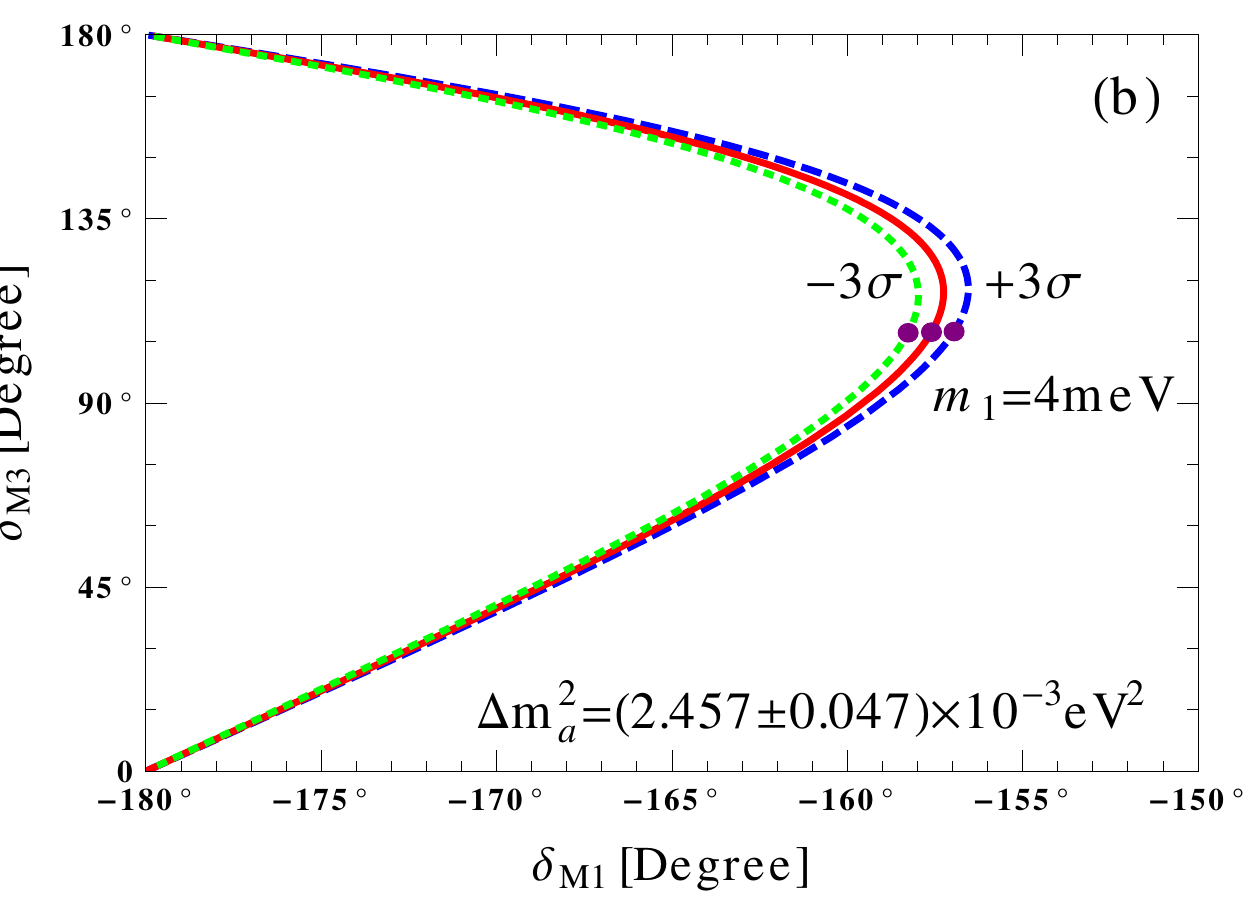}
\includegraphics[width=0.4\textwidth]{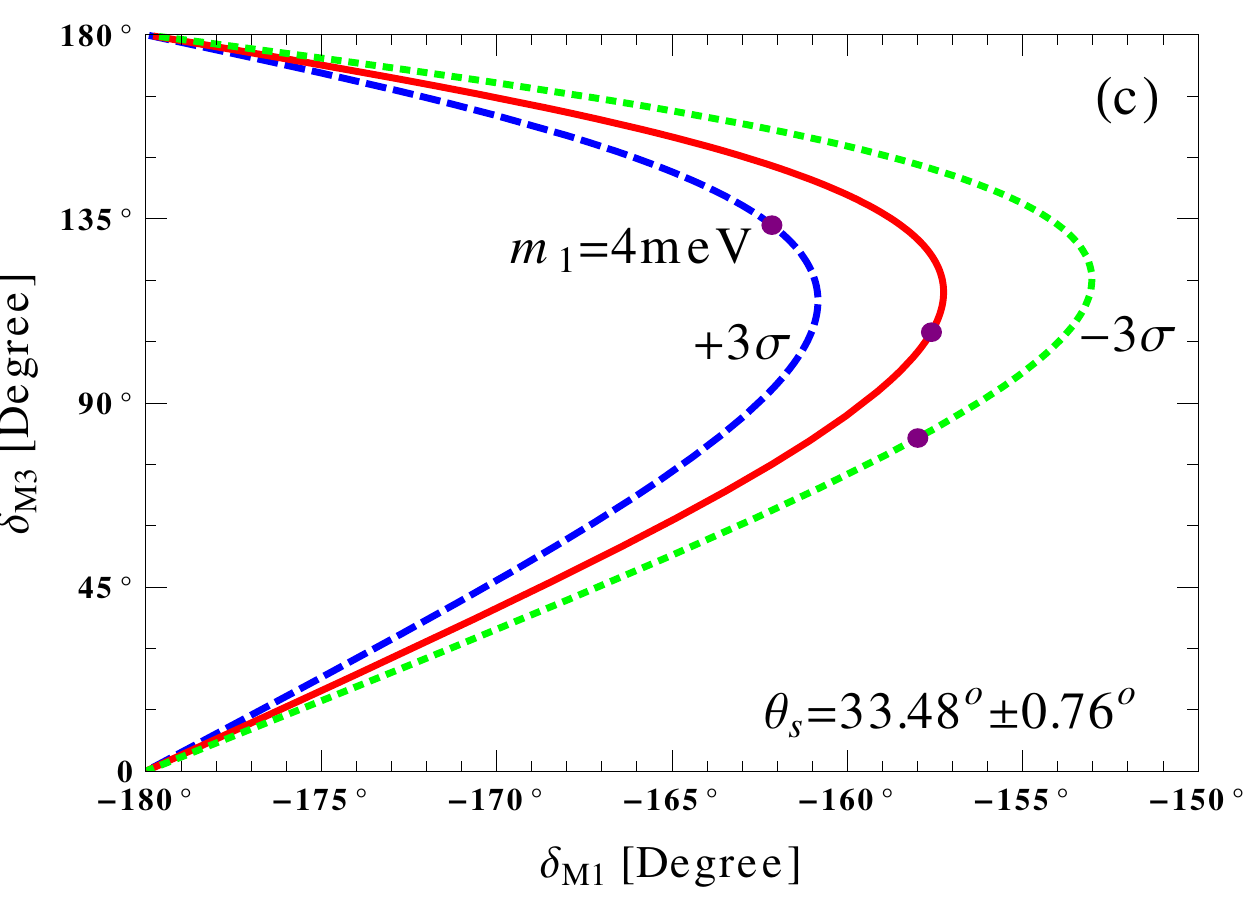}
\includegraphics[width=0.4\textwidth]{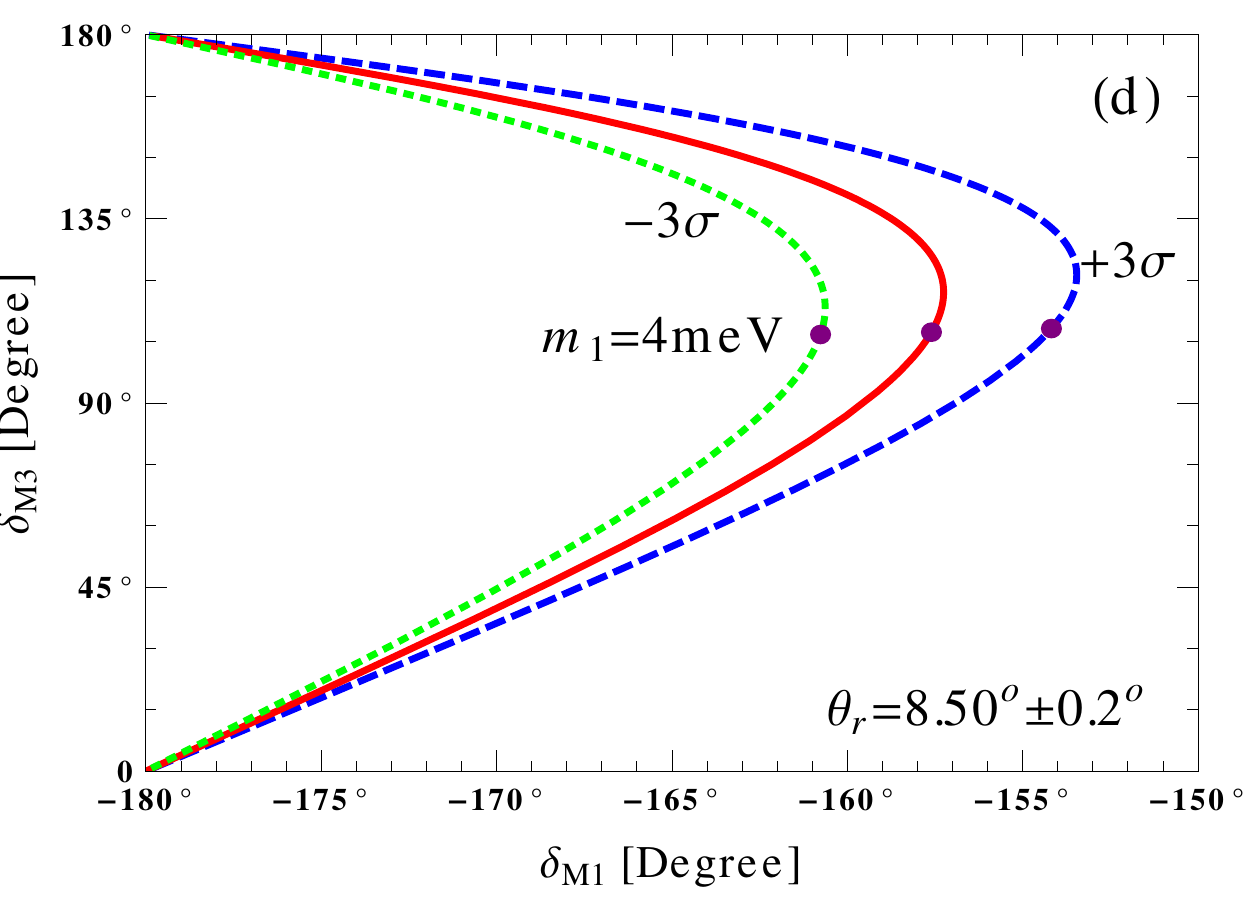}
\caption{The contamination from the $3\sigma$ uncertainty \cite{globalfit15} of
         oscillation parameters $\Delta m^2_s$, $\Delta m^2_a$, $\theta_r$,
         and $\theta_s$, respectively.}
\label{fig:dCPM13_3sigma}
\end{figure}
We show in \gfig{fig:dCPM13_3sigma} the $3\sigma$ variation of the predicted $\dMa(m_1)$ and $\dMc(m_1)$
on the four input oscillation parameters ($\Dma$, $\Dms$, $\Tr$, $\Ts$). The $\dMa$--$\dMc$ curve moves
to the left when increasing the values of the solar parameter $\Dms$ or $\Ts$ and to the right
for the atmospheric mass split $\Dma$ or the reactor angle $\Tr$. While $\Dma$ and $\Tr$
mainly affect $\dMa$, the solar parameters mainly change $\dMc$. Note that the x- and
y-axes in \gfig{fig:dCPM13_3sigma} have quite different scale. The x-axis with plotted
range $(-180^\circ, -150^\circ)$ is stretched by a factor of 6 than the y-axis which is
plotted with the range $(0^\circ, 180^\circ)$. Even with this magnification in the x-axis,
the variation in $\dMc$ when changing $\Dma$ and $\Tr$ is not visible. Although it becomes
sizable when varying $\Dms$ and $\Ts$, the variation in $\dMa$ is much smaller than in
$\dMc$. In addition, the variation from mass splits, $\Dms$ and $\Dma$, is relatively
smaller than the one from mixing angles, $\Ts$ and $\Tr$. Precisely measuring the oscillation
parameters ($\Dma$, $\Dms$, $\Tr$, $\Ts$), especially the two mixing angles, can
help to determine the Majorana CP phases from vanishing $\mee$.

The same thing happens for the lower limit of $\mee$ \cite{Dueck:2011hu}.
For inverted hierarchy (IH), the effective mass $\mee$ cannot vanish. When varying the
Majorana CP phases, $\mee$ spans a range. Its minimal value $\mee^{min}$ is a result
of minimizing $\mee$ with respect to $\dMa$ and $\dMc$. Consequently, $\mee^{min}$
is also independent of $\dMa$ and $\dMc$, but a function of the smallest mass $m_1$
and the four oscillation parameters ($\Dma$, $\Dms$, $\Tr$, $\Ts$). Similarly, the
largest uncertainty comes from the solar sector, especially $\Ts$. With the global fit
\cite{globalfit14} at that time, the $3\sigma$ uncertainty in $\Ts$ can introduce a
factor of 6 difference in the required target mass for given sensitivity \cite{Ge:2015bfa}. The only
difference is that for IH, the two circles in \gfig{fig:mee1} cannot touch each other since
$L_2 < L_1 - L_3$. In this situation, the minimal value $\mee^{min} = L_1 - L_2 - L_3$
happens at $\dMa = \pm 180^\circ$ and $\dMc = 0^\circ$. For NH, the minimal value $\mee^{min}$ can
touch down to zero if \geqn{eq:conditions} holds. Two conditions appear for the
real and imaginary parts of $\mee^{min}$ to eliminate two degrees of freedom
and produce the two equations in \geqn{eq:cos}.

As pointed out in \cite{Ge:2015bfa}, both reactor neutrino oscillation and $\nuless$
involve the same electron-electron channel. These two different phenomena share the
same set of oscillation parameters ($\Dma$, $\Dms$, $\Tr$, $\Ts$). The measurement at
reactor neutrino experiments can help to reduce the uncertainty in $\nuless$ decay
measurement. Since the reactor neutrino oscillation is well established by the observations
at Daya Bay \cite{DayaBay}, RENO \cite{RENO}, and Double Chooz \cite{DoubleChooz},
the precision measurement of oscillation parameters there can help reduce the uncertainty
in $\nuless$ decay,
especially when combining the measurements at both short and medium
baseline reactor experiments. The short baseline (Daya Bay, RENO, Double Chooz) can
measure the fast frequency oscillation due to $\Dma$ and $\Tr$ while the medium baseline
(such as JUNO \cite{JUNO} and RENO-50 \cite{RENO50}) has better resolution on the slow
frequency oscillation due to $\Dms$ and $\Ts$ \cite{Ge:2012wj}. Together, all of the four
oscillation parameters ($\Dma$, $\Dms$, $\Tr$, $\Ts$) can be measured precisely.
The advantage of reactor neutrino experiments is not just about measuring the smallest
mixing angle $\Tr$ and the neutrino mass hierarchy, but also significantly reducing
the uncertainty in $\nuless$ decay from oscillation parameters.

The effect of $\Tr$ and $\Ts$ uncertainties on $\nuless$ decay can be found in \cite{Lindner:2005kr}
and \cite{Dueck:2011hu}. With the reactor angle $\Tr$ being precisely measured \cite{DayaBayHistory, DoubleChoozHistory},
the major uncertainty now mainly comes from the solar angle $\Ts$ \cite{Dueck:2011hu, Ge:2015bfa}.
The next-generation of reactor neutrino experiments with medium baseline, such as JUNO \cite{JUNO}
and RENO-50 \cite{RENO50} experiments can have very precise measurement on $\Ts$, with relative
uncertainty down to $\sim 0.3\%$ \cite{JUNO,Ge:2012wj}. 
The combination of Daya Bay and JUNO, one short baseline and the other medium baseline,
can measure the four oscillation parameters $\Delta m^2_s$, $\Delta m^2_a$, $\theta_r$,
and $\theta_s$ very precisely.
\begin{figure}[h]
\centering
\includegraphics[height=0.45\textwidth,angle=-90]{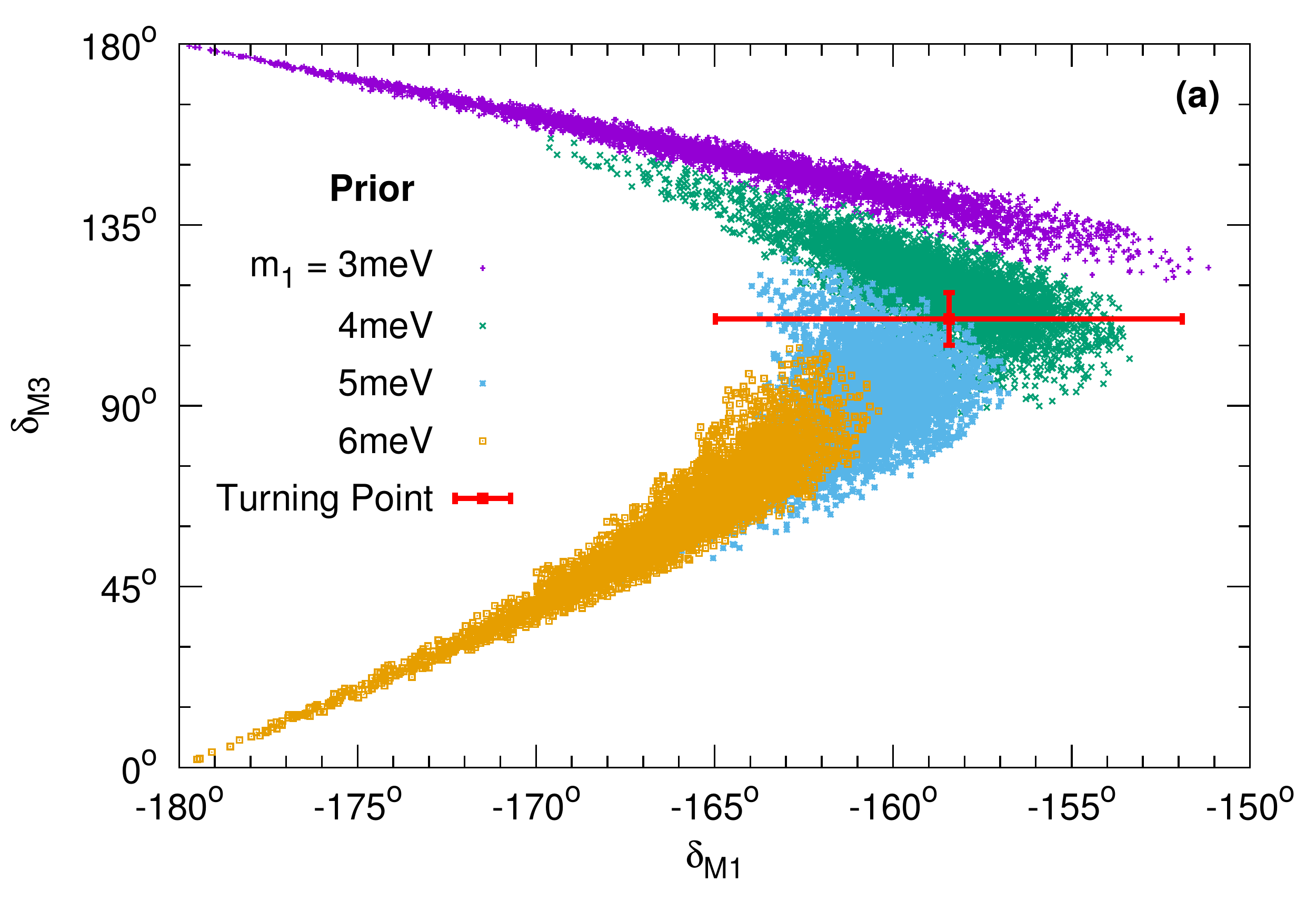}
\qquad
\includegraphics[height=0.45\textwidth,angle=-90]{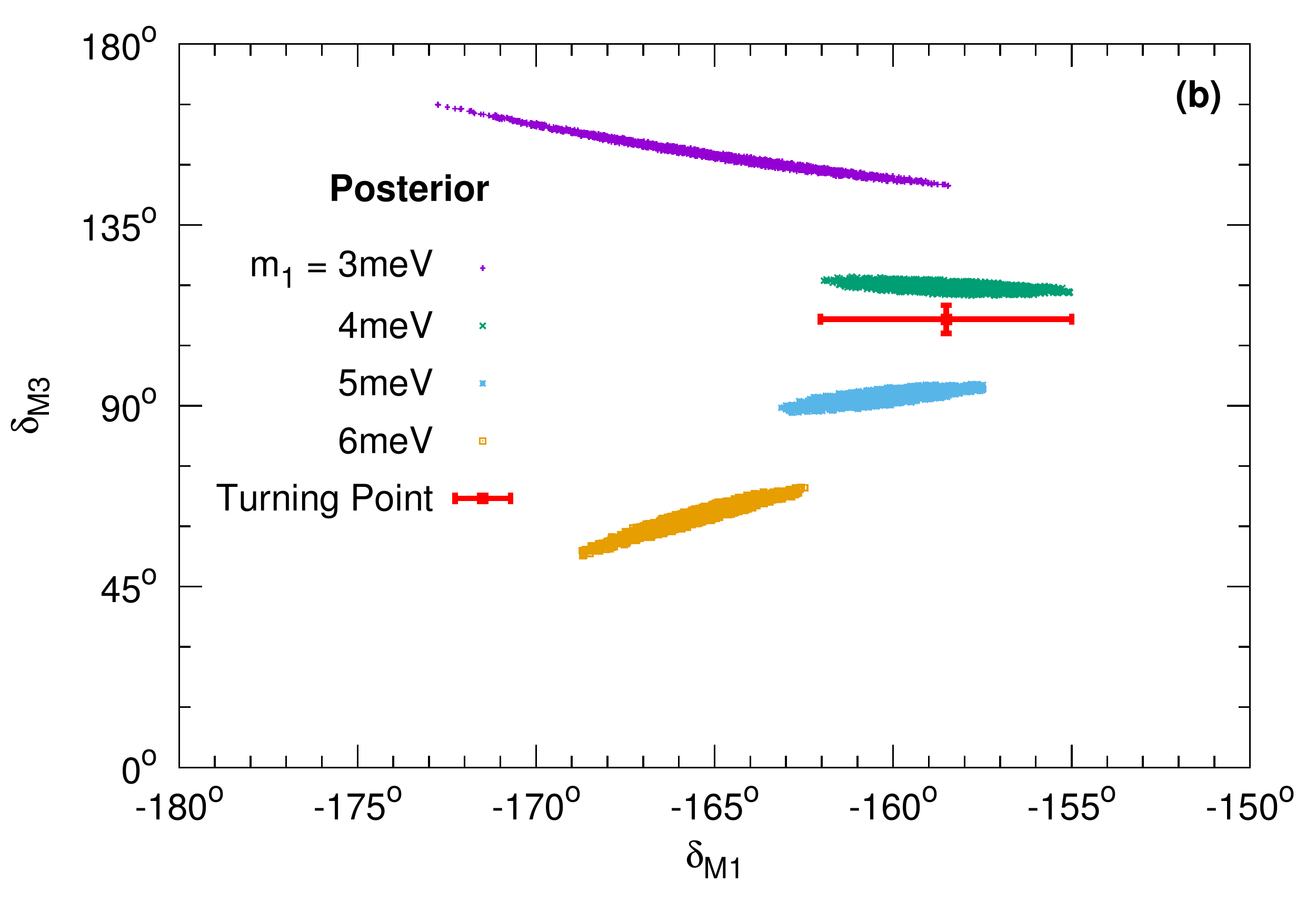}
\includegraphics[height=0.45\textwidth,angle=-90]{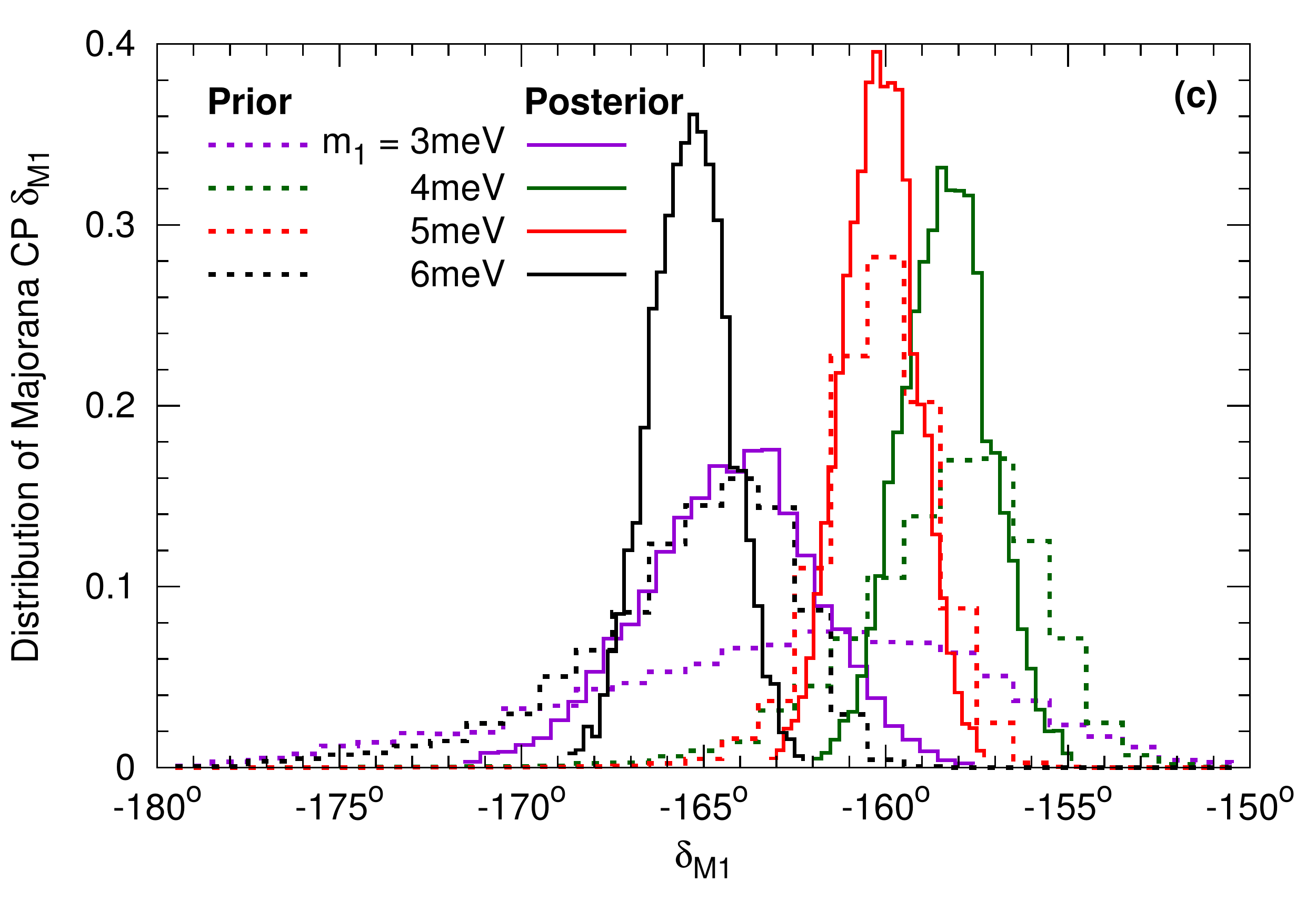}
\qquad
\includegraphics[height=0.45\textwidth,angle=-90]{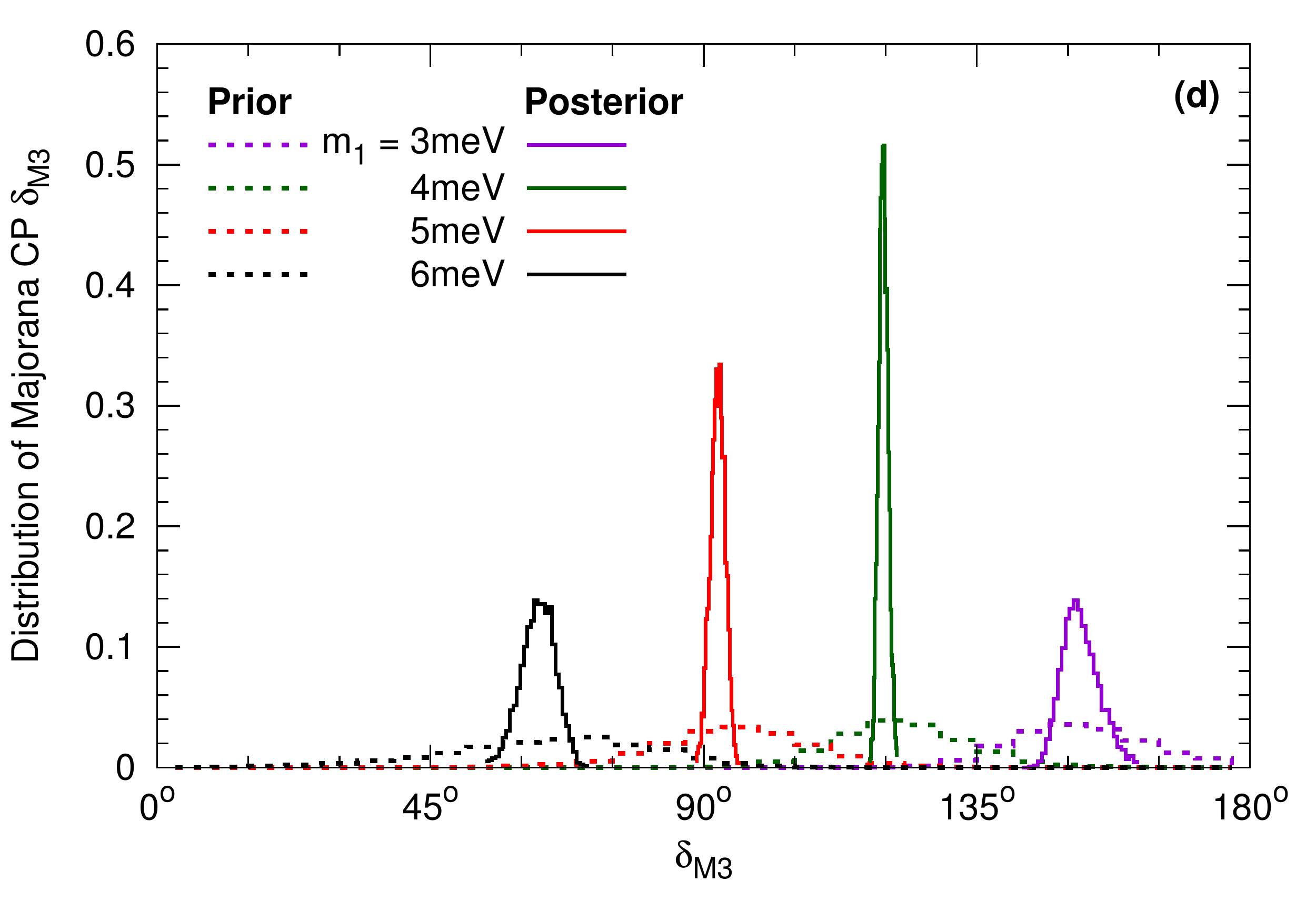}
\caption{The prior (before JUNO) and posterior (after JUNO) distributions of the
         Majorana CP phases $\dMa$ and $\dMc$ determined from the Majorana Triangle
         with $m_1 = 3,4,5,6\,\mbox{meV}$, respectively. In the subplots we show
         (a) the prior 2-dimensional distribution $\dMa$--$\dMc$ with $\chi^2 < 9$,
         (b) the posterior 2-dimensional distribution $\dMa$--$\dMc$ with $\chi^2 < 9$,
         (c) the 1-dimensional distribution of $\dMa$, and
         (d) the 1-dimensional distribution of $\dMc$. The red crosses in (a) and (b)
         indicate the $3 \sigma$ uncertainties in the values of $\dMa$ and $\dMc$ at turning points.}
\label{fig:nulessJUNO}
\end{figure}

We use NuPro \cite{NuPro} to simulate JUNO for illustration and generate scattered
points in the four-dimensional parameter space ($\Dma$, $\Dms$, $\Tr$, $\Ts$) with
help of the Bayesian Nested Sampling algorithm \cite{Bayesian}
implemented in MultiNest \cite{MultiNest}. Given a specific value of $m_1$, we obtain
the distribution of predicted Majorana CP phases $\dMa$ and $\dMc$ according to
\geqn{eq:cos}. The \gfig{fig:nulessJUNO} shows the results as both two-dimensional
scattered plots with $\chi^2 < 9$ and one-dimensional histograms for the whole parameter
space. As an illustration, we take four
typical values $m_1 = 3,4,5,6\,\mbox{meV}$ within the considered range \geqn{eq:conditions},
or equivalently $2.3\,\mbox{meV} \lesssim m_1 \lesssim 6.3\,\mbox{meV}$. With the
current global fit \cite{globalfit15} as prior constraints, the scattered points
for $m_1 = 3,4,5,6\,\mbox{meV}$ overlap with each other in \gfig{fig:nulessJUNO}\,(a).
In comparison, the posterior distributions in \gfig{fig:nulessJUNO}\,(b) after including
JUNO are well separated from each other. Especially, the predictions for $m_1 = 3\,\mbox{meV}$
and $m_1 = 6\,\mbox{meV}$ no longer connects with the trivial solutions $\dMa = -180^\circ$.
Of the two Majorana CP phases $\dMa$ and $\dMc$, whose marginalized probability distributions
are shown in \gfig{fig:nulessJUNO}\,(c) and \gfig{fig:nulessJUNO}\,(d), we observe more
significant reduction in the uncertainty in $\dMc$ than in $\dMa$. This feature is consistent
with the earlier observations that the uncertainty in $\Ts$ has larger effect in $\dMc$ than
in $\dMa$ and JUNO can mainly reduce the uncertainty of the solar parameters.

\begin{table}[h]
\centering
\setlength{\tabcolsep}{4mm}
\begin{tabular}{cc||cc|cc}
  \multicolumn{2}{c||}{\multirow{2}{*}{$1 \sigma$ Uncertainties}} & \multicolumn{2}{c|}{Prior} & \multicolumn{2}{c}{Posterior} \\
  & & $\dMa$ & $\dMc$ & $\dMa$ & $\dMc$ \\
\hline\hline
  \multirow{4}{*}{$m_1 = $}  
& $3\,\mbox{meV}$ & $-164^\circ \pm 5.4^\circ$ & $149^\circ \pm 10.8^\circ$ 
                  & $-165^\circ \pm 2.4^\circ$ & $152^\circ \pm  3.3^\circ$  \\
& $4\,\mbox{meV}$ & $-159^\circ \pm 2.6^\circ$ & $120^\circ \pm 10.8^\circ$ 
                  & $-158^\circ \pm 1.2^\circ$ & $120^\circ \pm  0.8^\circ$  \\
& $5\,\mbox{meV}$ & $-161^\circ \pm 1.5^\circ$ & $91.7^\circ \pm 12.5^\circ$
                  & $-160^\circ \pm 1.0^\circ$ & $92.2^\circ \pm  1.2^\circ$ \\
& $6\,\mbox{meV}$ & $-166^\circ \pm 3.1^\circ$ & $61.7^\circ \pm 17.1^\circ$
                  & $-166^\circ \pm 1.1^\circ$ & $62.4^\circ \pm  2.9^\circ$ \\
  \hline
  \multicolumn{2}{c||}{\multirow{2}{*}{Turning Point}}
                                 & \multicolumn{2}{c|}{$m_1 = (4.31 \pm 0.46)\,\mbox{meV}$}
                                 & \multicolumn{2}{c }{$m_1 = (4.29 \pm 0.05)\,\mbox{meV}$} \\
                               & & $-158^\circ \pm 2.2^\circ$ & $112^\circ \pm 2.2^\circ$
                                 & $-159^\circ \pm 1.2^\circ$ & $112^\circ \pm 1.2^\circ$
\end{tabular}
\caption{The $1\sigma$ uncertainties of the two Majorana CP phases $\dMa$ and $\dMc$, the turning
         point parameters ($m_1$, $\dMa$, $\dMc$), and the upper limits $\mee^{upper}$ in
         (\ref{eq:mee-upper}) before and after JUNO/RENO-50.}
\label{tab:nulessJUNO}
\end{table}

Since the scales of $\dMa$ and $\dMc$ as shown in \gfig{fig:nulessJUNO} are not the same,
we list their $1\sigma$ uncertainties in \gtab{tab:nulessJUNO}. The uncertainty of
$\dMa$ is reduced by a factor of around $1.5\sim 3$ while $\dMc$ by a factor of $3\sim 10$.
This reflects the fact that the reduced uncertainty depends mostly on the
solar parameters $\Dms$ and $\Ts$. The position uncertainty of the turning point even
reduces by a factor of $10$, from $0.46\,\mbox{meV}$ to $0.05\,\mbox{meV}$. On the other
hand, the uncertainties in the value $\dMa$ and $\dMb$ at the turning point reduce by
only a factor of 2. Note that $\dMa$ and $\dMb$ have the same uncertainty, since they
are correlated with each other, $\dMc = 270^\circ + \dMa$, at the turning point.
Altogether, given the smallest mass $m_1$, the Majorana Triangle with vanishing $\mee$
can predict the Majorana CP phases to degree level with the help of medium baseline
reactor neutrino experiment such as JUNO or RENO-50. At that time, the largest uncertainty
would almost entirely come from the unknown mass scale $m_1$ \cite{m1Uncertainty} and
$\nuless$ decay determination on $\mee$.

\section{Sensitivity to Majorana CP Phases}
\label{sec:sensitivity}

As demonstrated in \gsec{sec:triangle}, from a non-observation of $\nuless$ decay
we can still infer the Majorana CP phases $\dMa$ and $\dMc$. In practice, non-observation
can not lead to a exactly vanishing $\mee$, but an upper limit on it.
The inferred $\dMa$ and $\dMc$ from the Majorana Triangle inevitably have uncertainty
from the $\nuless$ decay measurement, even with precision measurement of the
oscillation parameters by reactor experiments. If the upper limit on $\mee$ is
too large, the possible solution of Majorana CP phases can scan the whole region
from the intersection $I_1$ to $I_2$ shown in \gfig{fig:mee1}.
In other words, the Majorana CP phases can cross the trivial values $0^\circ$ or
$\pm 180^\circ$. Requiring non-trivial solutions of the two Majorana
CP phases, would place an upper limit on the uncertainty of $\mee$ and hence
requirement on the design of future $\nuless$ experiments. 

\begin{figure}[h]
\centering
\includegraphics[scale=0.55]{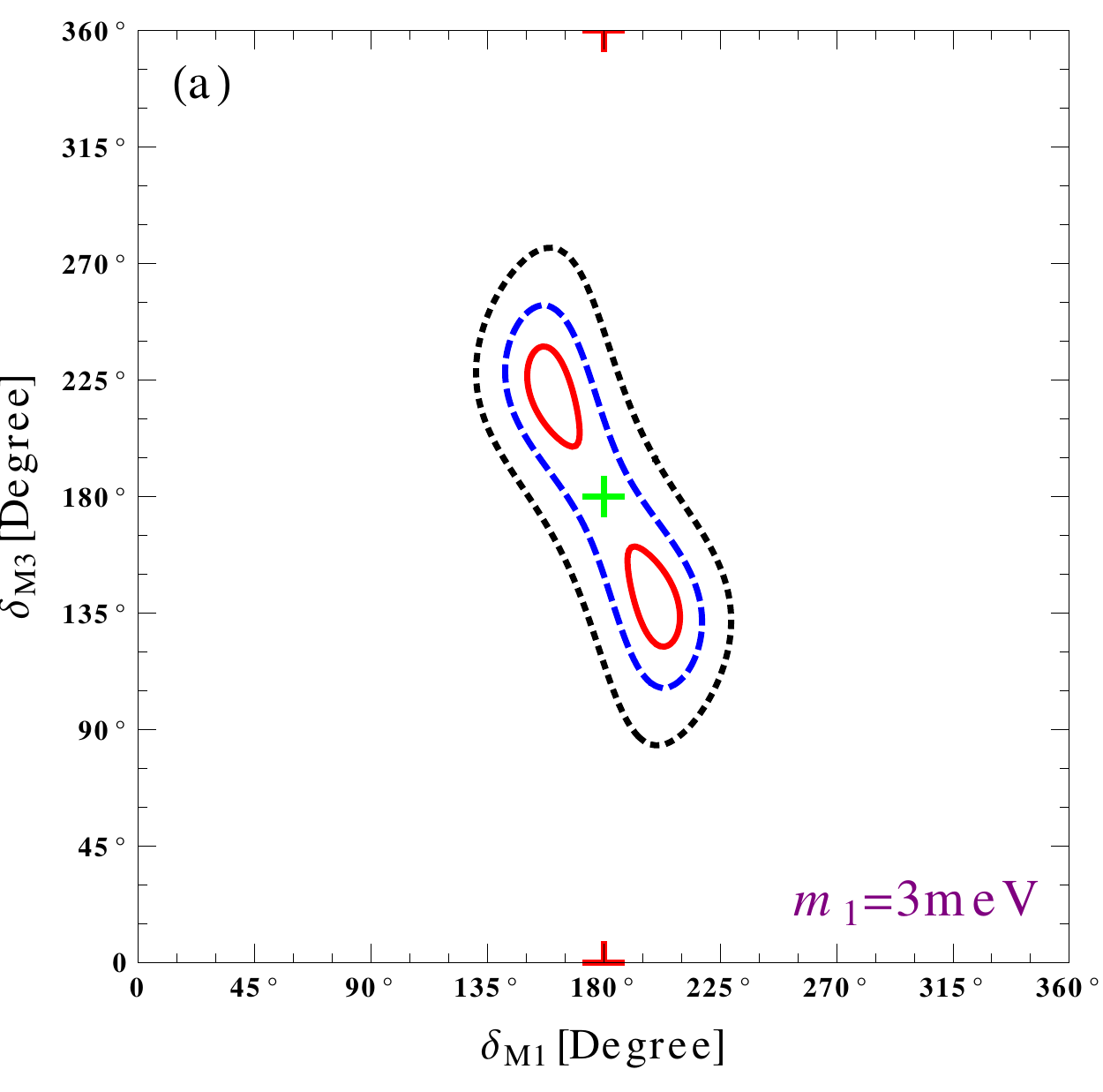}
\includegraphics[scale=0.55]{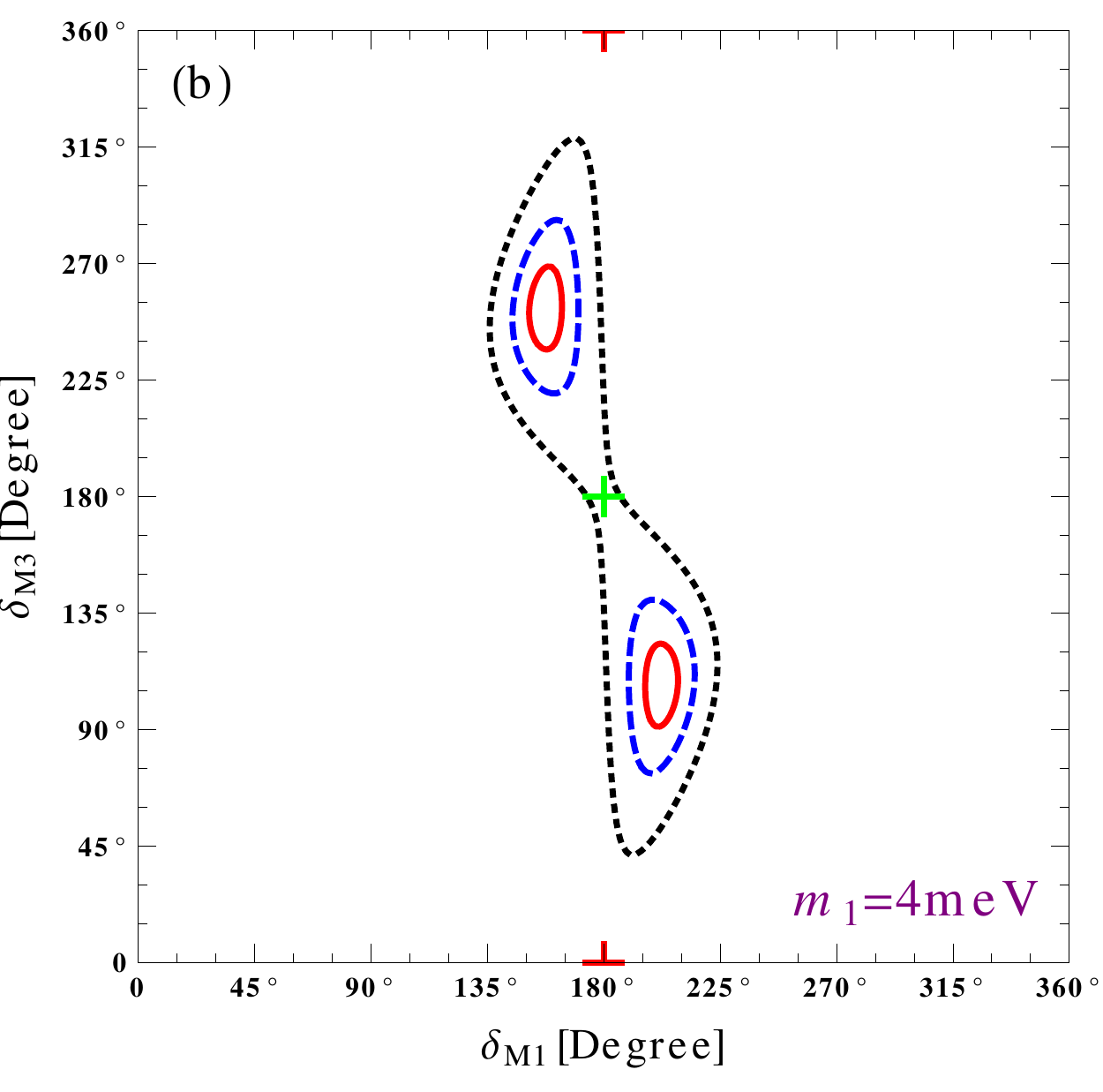}
\includegraphics[scale=0.55]{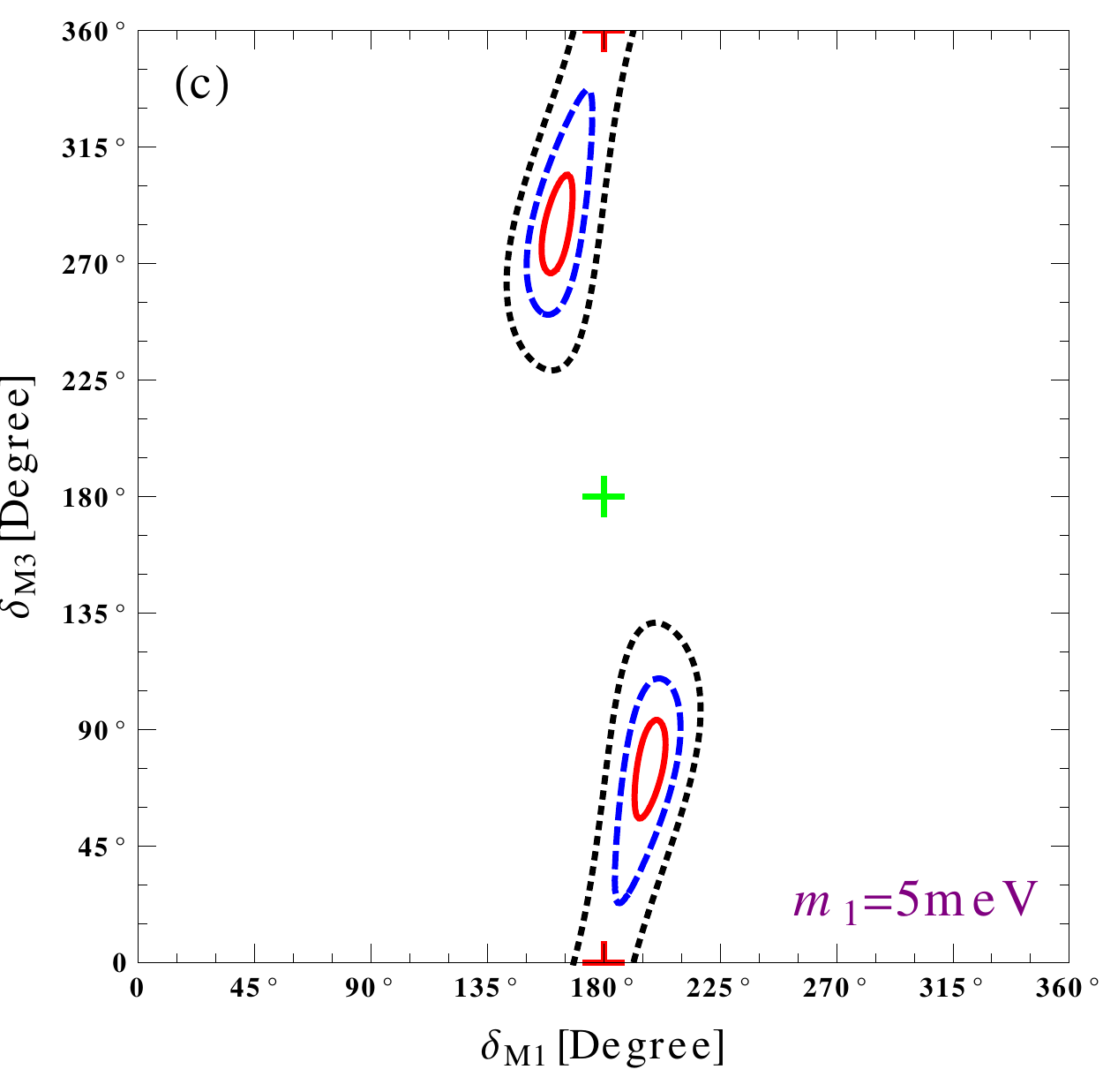}
\includegraphics[scale=0.55]{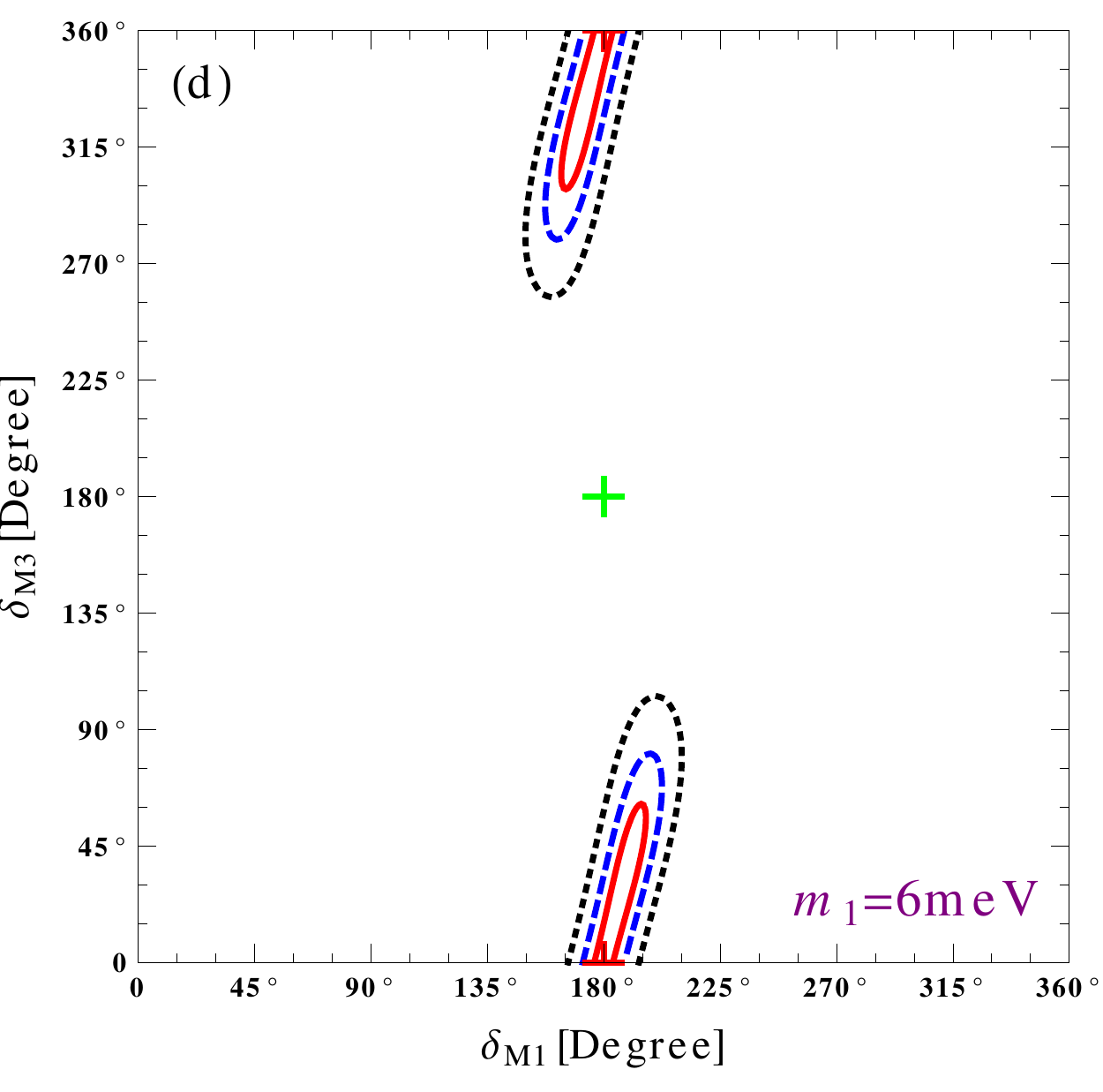}
\caption{The contour plot of $\mee$ in the $\dMa$--$\dMc$ space with
         the lightest mass eigenvalues $m_1 = (3,4,5,6)\,\mbox{meV}$ for the 4 panels
         and upper limits on $\mee = (0.3, 0.6, 1) \, \mbox{meV}$ depicted in 
         (\gred{\bf solid}, \gblue{\bf dashed}, {\bf dotted}) curves. In addition,
         the two trivial solutions ($\dMa = \dMc = 180^\circ$) and ($\dMa = 180^\circ$,
         $\dMc = 0^\circ, 360^\circ$) appear as green and red crosses, respectively.}
\label{fig:contours}
\end{figure}

We show in \gfig{fig:contours} the $\mee$ contour on the $\dMa$--$\dMc$ plane
for specific values of the smallest mass, $m_1 = (3,4,5,6)\,\mbox{meV}$ in the four subplots.
Each subplot shows three contours with $\mee = (0.3, 0.6, 1)\,\mbox{meV}$. Since we show
the full range of $\dMa$ and $\dMc$ from $0^\circ$ to $360^\circ$, we can see two non-trivial
solutions of vanishing $\mee$. The one in the lower-right quadrant, $180^\circ < \dMa < 360^\circ$
and $0^\circ < \dMc < 180^\circ$ corresponds to the solution shown in earlier plots while
there is a symmetric solution in the upper-left quadrant. In addition, there are two trivial
points of the Majorana CP phases, shown as green and red crosses in \gfig{fig:contours}.
The first, $\dMa = \dMc = 180^\circ$, happens for $L_2 - L_3 < L_1 < L_2$ while the second for
$\dMa = 180^\circ$ and $\dMc = 0^\circ$. The contours around the two non-trivial solutions
of vanishing $\mee$ would merge into a single contour if the trivial points $\dMa = 180^\circ$
and $\dMc = 0^\circ (360^\circ)$ are also covered for larger value of $\mee$. Otherwise,
the two solutions are isolated and non-trivial Majorana CP phases can be inferred.

\begin{figure}[h]
\centering
\includegraphics[scale=0.6]{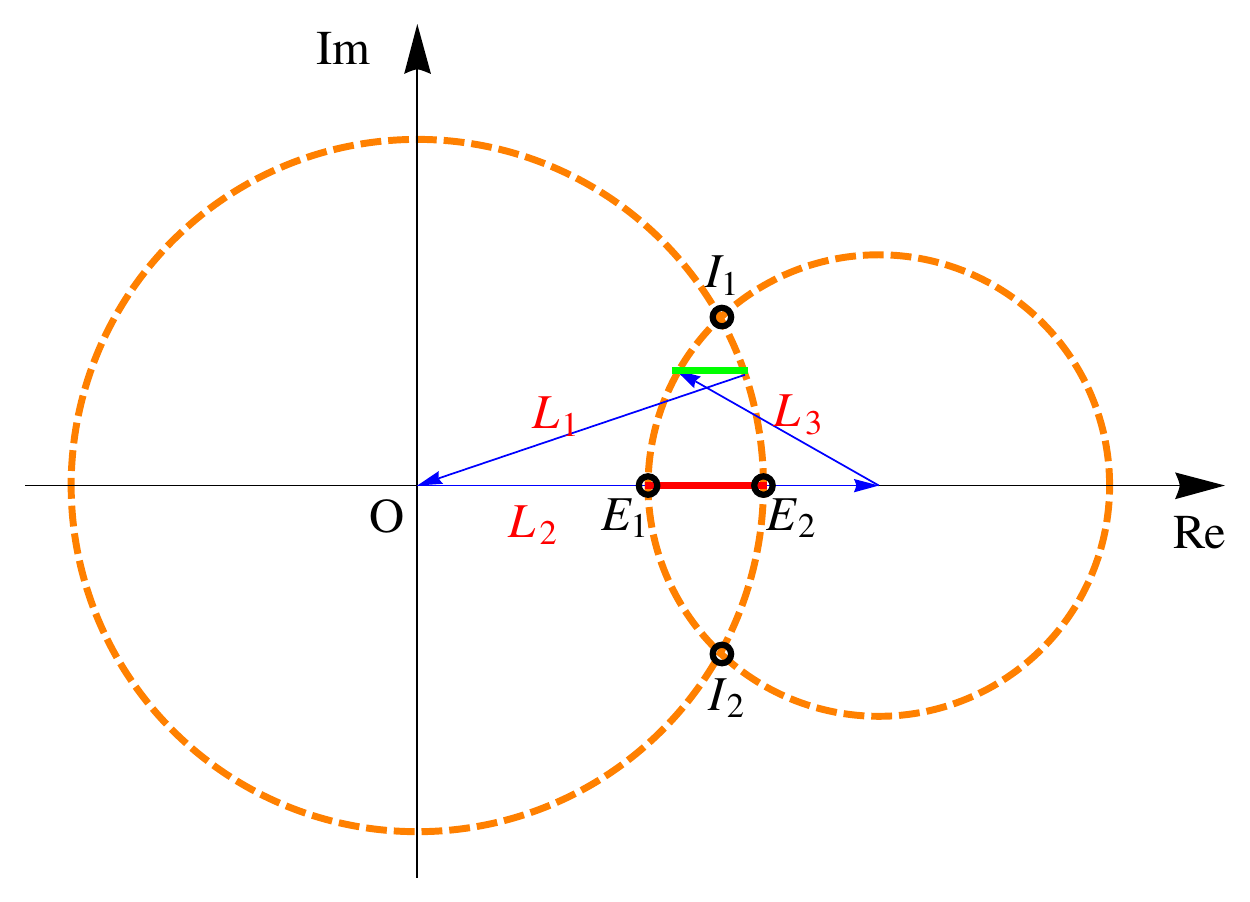}
\qquad
\includegraphics[scale=0.52]{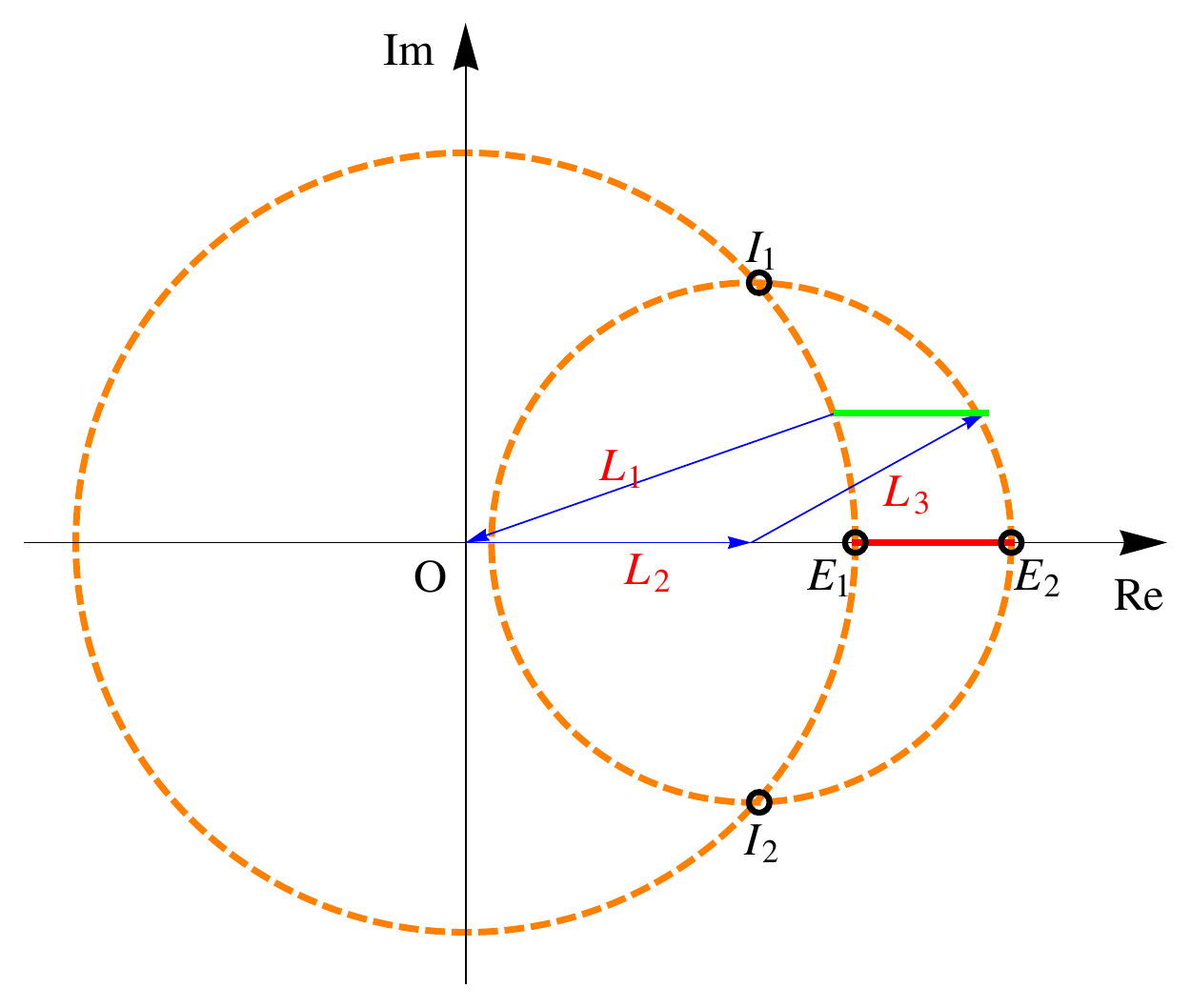}
\caption{Geometrical illustration of the required sensitivity (red bar), or equivalently
         upper limit $\mee^{upper}$, to have non-trivial solutions
         of the Majorana CP phases if $\nuless$ decay is not observed. The left plot is for
         $L_1 < L_2$ while the right for $L_1 > L_2$.}
\label{fig:mees}
\end{figure}
For illustration, we show the non-zero $\mee$ as a green bar in \gfig{fig:mees}.
Given the uncertainty $\Delta(\mee)$, the green bar can slip around the intersection points
$I_1$ or $I_2$, as long as $\mee \leq \Delta(\mee)$. The largest value of $\mee$ between
$I_1$ and $I_2$ is the distance between $E_1$ and $E_2$. If the green bar is longer than
the red bar between $E_1$ and $E_2$, it can cross the x-axis and lead to trivial solutions,
namely $\dMa = \pm 180^\circ$ and $\dMc = 0^\circ, 180^\circ$ when it lies on the x-axis.
To guarantee non-trivial Majorana CP phases, the sensitivity $\mee^{upper}$ cannot be larger
than the length of the red bar,
\begin{equation}
  \mee^{upper}
<
\begin{cases}
  L_1 + L_3 - L_2
\qquad \mbox{for} \qquad
  L_1 < L_2 \,,
\\
  L_2 + L_3 - L_1
\qquad \mbox{for} \qquad
  L_1 > L_2 \,.
\end{cases}
\label{eq:mee-upper}
\end{equation}

\begin{figure}[h]
\centering
\includegraphics[width=8cm]{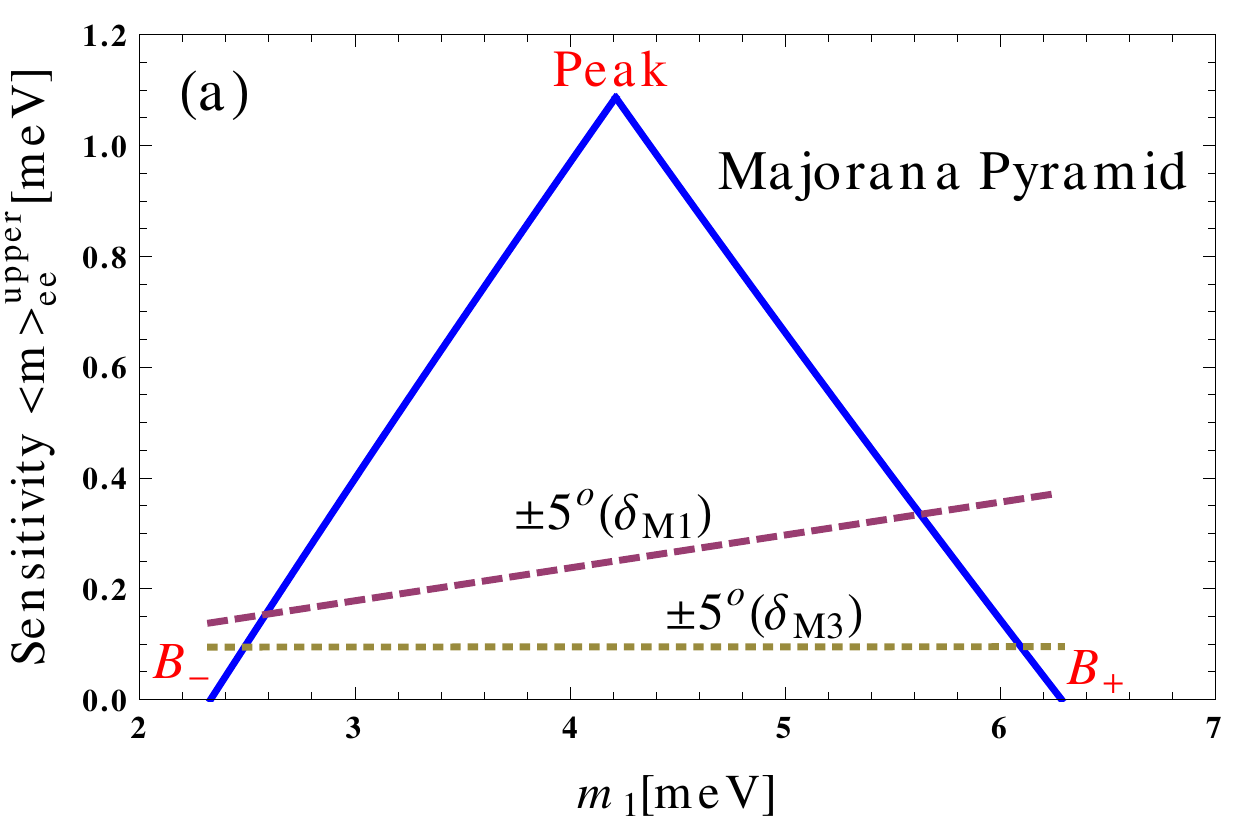}
\qquad
\includegraphics[width=8cm]{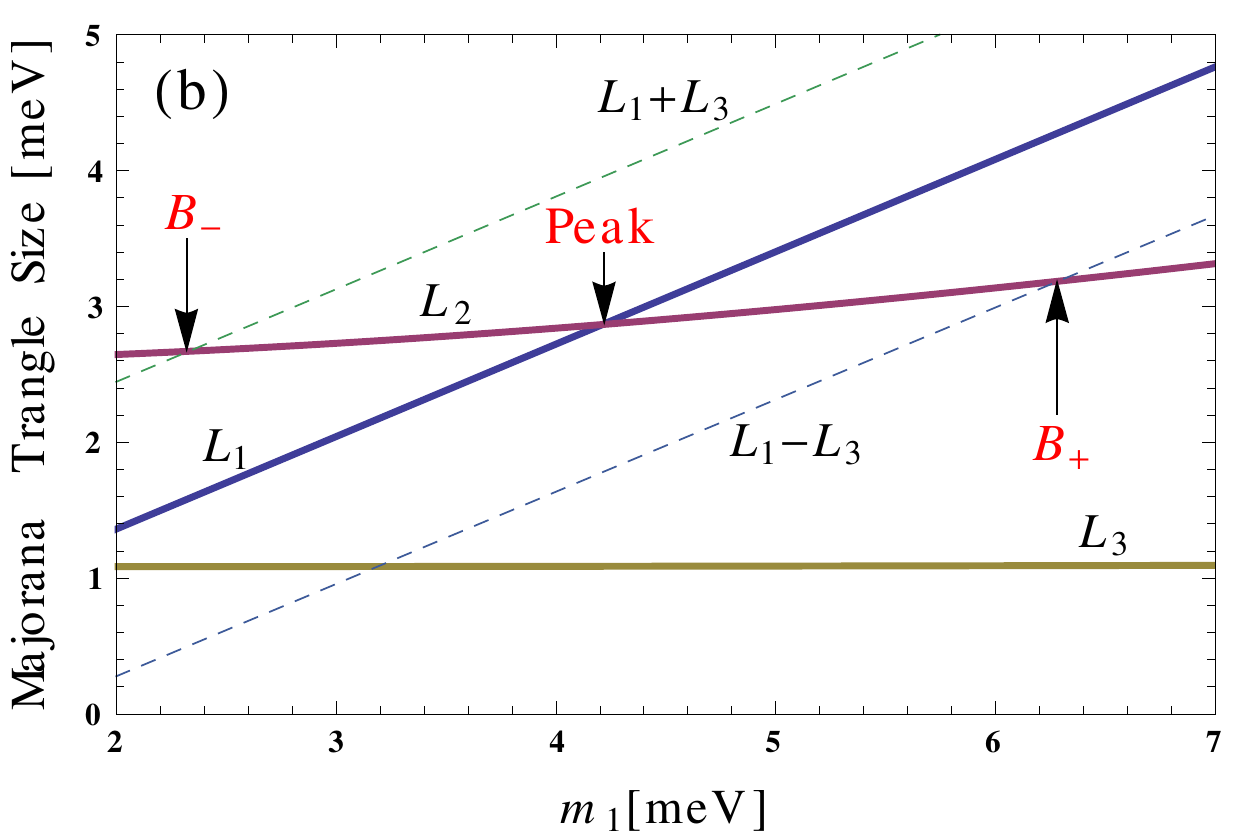}
\caption{(a) The required sensitivity $\mee^{upper}$ on $\mee$ to guarantee non-trivial solutions
         of the Majorana CP phases, and
         (b) the corresponding boundary and peak points.}
\label{fig:meeUpLimit}
\end{figure}

In \gfig{fig:meeUpLimit}\,(a) we show the required sensitivity $\mee^{upper}$ to guarantee
non-trivial Majorana CP phases as a function of the smallest mass $m_1$.
Its shape resembles a pyramid, leading to a metaphor
that the two Majorana CP phases $\dMa$ and $\dMc$ hiding in the {\it Majorana Pyramid} as
snails lingering around as long as the sensitivity $\mee^{upper}$ is not low enough to
touch them. The peak appears in the middle when $L_1 = L_2$ with height being $L_3$,
\begin{equation}
  \mee^{upper}
\leq
  L_3
=
  s^2_r \sqrt{\frac {s^4_s}{c^2_s - s^2_s} \Dms + \Dma}
\approx
  s^2_r \sqrt{\Dma}
\approx
  1.1 \, \mbox{meV} \,.
\label{eq:meeUpLimit}
\end{equation}
It is interesting to see that the peak appears at $m^{peak}_1 \equiv s^2_s \sqrt{\Dms/(c^2_s - s^2_s)}$
which is around the turning point \geqn{eq:turningPoint}.
While the peak position is determined by the solar parameters $\Dms$ and $\Ts$, its height
mainly is a function of the atmospheric mass split $\Dma$ and the reactor angle $\Tr$.
From the top of the Majorana Pyramid, the sensitivity $\mee^{upper}$ decreases linearly
with the deviation $m_1 - m^{peak}_1$ from the peak position and vanishes at the two boundaries
$B_\pm$ ($L_1 - L_2 = \pm L_3$) that corresponding to $m_1 = 2.3\,\mbox{meV}$ and $m_1 = 6.3\,\mbox{meV}$,
respectively. Both peak and boundaries are functions of only the oscillation parameters
($\Dma$, $\Dms$, $\Tr$, $\Ts$) and are independent of any other unmeasured parameters.
The Majorana Pyramid is well defined, especially after JUNO/RENO-50. To make the
picture explicit, we show in \gfig{fig:meeUpLimit}\,(b) the three sides ($L_1$, $L_2$, $L_3$)
of the Majorana Triangle as functions of the smallest mass $m_1$ and indicate their relations
with the peak and boundary positions. While the peak happens at the crossing point of
$L_1$ and $L_2$, the boundaries $B_\pm$ happens at the crossing point of $L_2$ and $L_1 \mp L_3$.

\begin{table}[h]
\centering
\setlength{\tabcolsep}{4mm}
\begin{tabular}{c||cccc}
 \multirow{2}{*}{$\mee^{upper}$ (meV)} & \multicolumn{4}{c}{The smallest mass $m_1$} \\ 
& 3 meV & 4 meV & 5 meV & 6 meV \\
\hline \hline
Prior     & $0.23 \pm 0.18$ & $0.75 \pm 0.21$ & $0.90 \pm 0.18$ & $0.48 \pm 0.26$ \\
Posterior & $0.21 \pm 0.06$ & $0.75 \pm 0.06$ & $0.97 \pm 0.06$ & $0.48 \pm 0.06$
\end{tabular}
\caption{The $1\sigma$ uncertainty of the required sensitivity $\mee^{upper}$ of the $\nuless$ decay
         measurement for extracting non-trivial Majorana CP phases, before (Prior) and after
         (Posterior) JUNO/RENO-50.}
\label{tab:meeUpLimit}
\end{table}

The sensitivity $\mee^{upper}$ also suffers from the uncertainty in solar parameters.
Although the largest value of $\mee^{upper}$ on the top of the Majorana Pyramid is
mainly a function of the atmospheric mass split $\Dma$ and the reactor angle $\Tr$,
see \geqn{eq:meeUpLimit}, the solar parameters $\Dms$ and $\Ts$ can still affect the
sensitivity. This is especially true for the parameter space off the peak. We list
the uncertainty of $\mee^{upper}$ for typical values $m_1 = 3,4,5,6\,\mbox{meV}$ in
\gtab{tab:meeUpLimit}. The uncertainty for $m_1 = 3,6\,\mbox{meV}$ is relatively
larger than for $m_1 = 4,5\,\mbox{meV}$. Without medium baseline reactor experiment
JUNO/RENO-50, the $3\sigma$ uncertainty at the peak can be as large as roughly $100\%$.
This can lead to a factor of $16$ difference in the required target mass for given
sensitivity \cite{Dueck:2011hu}. The JUNO/RENO-50 experiment can help to reduce this
uncertainty by a factor of $3.5$. Correspondingly, the uncertainty in the required
target mass reduces to around a factor of $2$. To guarantee the same sensitivity,
the detector size can be reduced by a factor of 8 when designing future $\nuless$
decay experiments.

In \gfig{fig:meeUpLimit} we also show how a changing $\dMa$ or $\dMc$ alone can affect
the effective mass $\mee$ for comparison. Around the vanishing $\mee$ we perturb
the Majorana CP phases by $5$ degrees and plot the value of non-zero $\mee$ as a function
of the smallest mass $m_1$. In other words, when the sensitivity on $\mee$ can be further
pushed to these two lines, we can not only infer non-trivial values of $\dMa$ and $\dMc$,
but constrain them with an uncertainty of only $5$ degrees. Different from the sensitivity
curve as Majorana Pyramid, the $\pm 5^\circ$ curves do not change much across the range of
$2.3\,\mbox{meV} \leq m_1 \leq 6.3\,\mbox{meV}$. They are much lower than the peak value
of $1.1\,\mbox{meV}$ and lies in the range $(0.1\sim0.4)\,\mbox{meV}$. Pinning down the
value of $\dMa$ and $\dMc$ is much harder than excluding trivial values, as expected.

\section{Conclusions}
\label{sec:conclusion}

In this paper we explore what a non-observation of $\nuless$ decay can teach us if we
assume that neutrinos are still Majorana particles.
Although the absence of $\nuless$ decay signal cannot verify the Majorana nature of neutrinos,
it provides the possibility of uniquely fixing the two Majorana CP phases simultaneously
from the Majorana Triangle with vanishing $\mee$. From the perspective of constraining model building,
this situation would be even better than measuring a nonzero $\mee$ which can
fix only one degree of freedom as a combination of the two Majorana CP phases. In addition,
the smallest mass eigenvalue is limited to a narrow window,
$2.3\,\mbox{meV} \lesssim m_1 \lesssim 6.3\,\mbox{meV}$. The medium baseline reactor
neutrino experiment JUNO/RENO-50 can help to significantly reduce the uncertainty in the
predicted Majorana CP phases. In addition, the uncertainty
in the required sensitivity for inferring non-trivial Majorana CP phases can also be reduced
by the precision measurement of solar parameters $\Dms$ and $\Ts$ at JUNO/RENO-50.

To guarantee the ability of identifying non-trivial Majorana CP phases, the $\nuless$ decay
experiment needs to touch the Majorana Pyramid with impressive sensitivity 
$\mee^{upper} \lesssim 1.1\,\mbox{meV}$. This sensitivity is roughly $10$ times smaller than
the ability of the next-generation $\nuless$ decay experiments which can touch down
to around $10\,\mbox{meV}$ and rough testify/falsify IH. Correspondingly, the detector scales
with $\mee^4$ and needs to expand by a factor of $10^4$ which seems like a mission impossible.
The situation may change if the background rate can be significantly suppressed below
the signal rate. Then, the detector only needs to scale with $\mee^2$ and expand by a
factor of $100$.


\end{document}